\documentclass[fleqn,usenatbib]{mnras}

\usepackage{newtxtext,newtxmath}
\usepackage[T1]{fontenc}

\DeclareRobustCommand{\VAN}[3]{#2}
\let\VANthebibliography\thebibliography
\def\thebibliography{\DeclareRobustCommand{\VAN}[3]{##3}\VANthebibliography}

\usepackage{graphicx}	
\usepackage{amsmath}	
\usepackage{siunitx}
\usepackage{pifont}

\title[PRISM]{PRISM: A  Non-Equilibrium, Multiphase Interstellar Medium Model for Radiation Hydrodynamics Simulations of Galaxies}

\author[H. Katz et al.] {Harley Katz,$^{1}$\thanks{E-mail:
  \href{mailto:harley.katz@physics.ox.ac.uk}{harley.katz@physics.ox.ac.uk}} Shenghua Liu,$^1$ Taysun Kimm,$^2$ Martin P. Rey,$^1$ Eric P. Andersson,$^3$ Alex J. Cameron,$^1$  \newauthor Francisco Rodriguez-Montero,$^1$ Oscar Agertz,$^3$ Julien Devriendt$^1$ and Adrianne Slyz$^1$
  \\
  $^1$Sub-department of Astrophysics, University of Oxford,
   Keble Road, Oxford OX1 3RH, United Kingdom \\
  $^2$Department of Astronomy, Yonsei University, 50 Yonsei-ro,
  Seodaemun-gu, Seoul 03722, Republic of Korea \\  
  $^3$Department of Astronomy and Theoretical Physics, Lund Observatory, Box 43, SE-221 00 Lund, Sweden
  }

\date{Accepted XXX. Received YYY; in original form ZZZ}

\pubyear{2022}

\begin{document}
\label{firstpage}
\pagerange{\pageref{firstpage}--\pageref{lastpage}}
\maketitle

\begin{abstract}
We introduce the {\small PRISM} interstellar medium (ISM) model for thermochemistry and its implementation in the {\small RAMSES-RTZ} code. The model includes a non-equilibrium primordial, metal, and molecular chemistry network for 115 species coupled to on-the-fly multifrequency radiation transport. {\small PRISM} accurately accounts for the dominant ISM cooling and heating processes in the low-density regime (i.e. $\rho<10^5\ {\rm cm^{-3}}$), including photoheating, photoelectric heating, H$_2$ heating/cooling, cosmic-ray heating, H/He cooling, metal-line cooling, CO cooling, and dust cooling (recombination and gas-grain collisions). We validate the model by comparing 1D equilibrium simulations across six dex in metallicity to existing 1D ISM models in the literature. We apply {\small PRISM} to high-resolution (4.5~pc) isolated dwarf galaxy simulations that include state-of-the-art models for star formation and stellar feedback to take an inventory of which cooling and heating processes dominate each different gas phase of a galaxy and to understand the importance of non-equilibrium effects. We show that most of the ISM gas is either close to thermal equilibrium or exhibits a slight cooling instability, while from a chemical perspective, the non-equilibrium electron fraction is often more than three times higher or lower than the equilibrium value, which impacts cooling, heating, and observable emission lines. Electron enhancements are attributed to recombination lags while deficits are shown to be due to rapid cosmic-ray heating. The {\small PRISM} model and its coupling to {\small RAMSES-RTZ} is applicable to a wide variety of astrophysical scenarios, from cosmological simulations to isolated giant molecular clouds, and is particularly useful for understanding how changes to ISM physics impact observable quantities such as metallic emission lines.
\end{abstract}

\begin{keywords}
ISM: general -- ISM: evolution -- galaxies: ISM --  galaxies: astrochemistry -- galaxies: evolution
\end{keywords}



\section{Introduction}
The physical state of the interstellar medium (ISM) of galaxies results from a rich set of interacting thermodynamic, radiative, and chemical processes \citep[e.g.][]{Draine2011}. Understanding these processes is of key importance for inferring the physical properties of galaxies (e.g. gas density, metallicity, ionization parameter) from various types of observations across cosmic time and for modelling galaxy formation from first principles.

In the low-redshift Universe, the ISM is predicted \citep[e.g.][]{Field1969,Mckee1977,Draine2011,Kim2017} to be a multiphase medium (i.e. multiple thermally stable gas phases coexist at the same pressure). While the ISM was originally envisioned to be a two-phase medium \citep{Field1969}, composed of a cold neutral medium (CNM) and a warm medium that can be both neutral (WNM) or ionized (WIM), it was later realized that the inclusion of supernova (SN) feedback generates a three-phase medium \citep{Cox1974,Mckee1977}: a CNM, a WNM/WIM, and a hot ionized medium (HIM). The exact properties of these phases (e.g. volume filling factors, total mass, temperature, etc.) are sensitive to the detailed balance of cooling and heating processes, which are themselves dependent on the chemical state and radiation field of the medium, as well as the star formation rate and feedback processes that inject additional energy and drive shocks/turbulence. These processes are well studied (both observationally and theoretically) for the ISM of the Milky Way \citep[e.g.][]{Wolfire1995,Wolfire2003} and theoretical models have provided additional insight into lower metallicity environments \citep[e.g.][]{Omukai2000,Omukai2005,Bialy2019}.

Due to the coupling between the chemistry and thermodynamics, different chemical species trace different phases of the ISM. For example, because O~{\small I} has the same ionisation potential as H~{\small I}, [O~{\small I}] 63$\mu$m observations primarily trace neutral gas while the [O~{\small II}]~$\lambda\lambda$3727 nebular line traces H~{\small II} regions. In principle, moving up the O ionization ladder will trace progressively hotter gas. A key open question is how to convert between the emission or absorption that we see and the underlying physical properties of the galaxy. 

1D spectral synthesis codes \citep[e.g.][]{Hjellming1966,Williams1967,Binette1985,Sutherland1993,Ferland1998} are commonly employed for this purpose. Grids of models can be constructed to infer global galactic properties \citep[][]{Kewley2001,Morisset2015} as well as combined to infer the underlying distribution of properties from a single set of spectral lines \citep[e.g.][]{Lebouteiller2022}. Some of the primary advantages of these models are that they are fast to execute computationally, which allows for the exploration of large regions of parameter space and the relevant spatial scales (e.g. the different ionization fronts) can be resolved. However, one of the disadvantages is that they, being 1D, cannot generally capture the complex and dynamic structure of the ISM along with all of the relevant, time-varying star formation and feedback.

For this reason, a complementary approach is to simulate the physics of the ISM using 3D models. Numerically modelling the ISM requires high spatial resolution to capture many of the important processes, on the scale of parsecs \citep[e.g.][]{Kim2017,Peters2017}, which is beyond the reach of most large-scale cosmological simulations \citep[][]{Vogelsberger2014,Dubois2014,Schaye2015}. 3D simulations of individual patches of the ISM can achieve the required spatial resolution and additionally model the complex interplay of star formation and feedback \citep[e.g.][]{Walch2015,Girichidis2016,Kim2017,Iffrig2017,Hill2018,Kim2020}, but they miss out on physics from larger scales and must put in dynamics (e.g. shearing) by hand. Simulations of isolated galaxies that also model much of the relevant physics are also now becoming possible \citep[e.g.][]{Kannan2020,Gutke2021,Richings2022,Andersson2022} and capture both large-scale galaxy motions and fluctuations in the radiation field from far away star formation. Cosmological zoom simulations of individual dwarf galaxies can now reach the required resolutions for ISM physics \citep[e.g.][]{Rey2019} and are able to capture further important physics such as accretion, a realistic circumgalactic medium, and environmental effects. However, the modelling techniques and the included physics varies widely between these different simulations. Importantly, to date, most simulations do not include any form of non-equilibrium metal chemistry.

In metal-enriched environments, it is often the case that metallic species dominate the cooling rate \citep[e.g.][]{Sutherland1993,Wiersma2009}. Thus, if a species is out of equilibrium, it can not only impact the observable emission, but also the thermodynamics of the gas. For the particular use case of comparing with spectroscopic observations, capturing both the correct evolution of the metal chemistry and the temperature state of the ISM is paramount. However, solving a large chemical network within a 3D simulation can be very computationally expensive. It is therefore common to ``post-process'' numerical simulations with photoionization codes or chemical networks in order to measure the metal ionization states and their relevant emission after the simulation has been run \citep[e.g.][]{Karen2018,Katz2019,Lupi2020,Gong2020,Barrow2020,Pallottini2022}. Simulations can also be run with substantially reduced chemical networks that target an atom or molecule of interest for a specific problem. One of the crucial assumptions that is often made when post-processing simulation outputs or when using a substantially reduced chemical network is that the atoms and molecules that are not followed self-consistently are in equilibrium. Depending on environment this assumption is not guaranteed to be true because the timescales of different physical processes (e.g. cooling and recombination) are not necessarily synchronized \citep[e.g.][]{Kafatos1973,Gnat2007}. 

There are now numerous 3D numerical simulations that employ much larger chemical networks \citep[e.g.][]{Glover2012a,Oppenheimer2013,Richings2014,Gray2015,Oppenheimer2018,Richings2022}. Due to the expense of solving these networks, few of these codes couple the chemistry to on-the-fly radiation hydrodynamics. However, within the same galaxy, the photoionization and photoheating rates can vary by many orders of magnitude \citep[e.g.][]{Rosdahl2015}, which can lead to strong local variations in ion abundances. Recently, \cite{Richings2022} applied the {\small LEBRON} method for approximate radiative transfer \citep{Hopkins2018} and found order of magnitude variations in the H and H$_2$ abundances in particular regions of temperature-density phase-space and changes of up to a factor of two in various emission line luminosities when comparing equilibrium and non-equilibrium solutions. Similarly, \cite{RTZ} showed that common bright emission line luminosities (e.g. [O~{\small III}]~$\lambda5007$, [O~{\small II}]~$\lambda\lambda3727$) calculated with various post-processing methods can be an order of magnitude different from the non-equilibrium value. 

In general, the impact of non-equilibrium physics on galaxy formation, especially that involving metal chemistry, remains relatively unexplored. This motivated the development of the {\small RAMSES-RTZ} code \citep{RTZ}, which leverages the computational efficiency of both the M1 radiative transfer scheme \citep{Levermore1984,Rosdahl2013} and a fast ODE solver inspired by \cite{Anninos1997}, as well as the parallel performance of the {\small RAMSES} code \citep{Teyssier2002} to accelerate the computation of non-equilibrium metal chemistry coupled to on-the-fly radiation hydrodynamics. The accelerations make it possible to run full cosmological simulations with such non-equilibrium physics. These first works with the {\small RAMSES-RTZ} code focused primarily on simulations of the early Universe and lower-metallicity environments \citep{Katz2022-popiii}. In this work, we introduce the {\small PRISM} model which computes the dominant heating and cooling processes in the ISM at both low and high metallicities. Our goal is to use high-resolution simulations of isolated galaxies to obtain a census of the relevant cooling and heating processes and to understand where and why ISM gas is in or out of thermal or chemical equilibrium. This framework presented here has already been employed to estimate the impact of temperature fluctuations on the mass-metallicity relation \citep{Cameron2022} and will form a baseline for future cosmological, isolated galaxy, and giant molecular cloud simulations.

This paper is organized as follows. In Section~\ref{chem_therm} we describe the chemical network and the various cooling and heating processes included in the {\small PRISM} model, comparing to equilibrium predictions of 1D models of the ISM in the literature. In Section~\ref{sims} we employ the {\small PRISM} model in high-resolution 3D simulations of two isolated dwarf galaxies and provide a census of the processes that contribute to cooling or heating the ISM and discuss the impact of non-equilibrium physics. Finally, in Sections~\ref{sec:limits} and \ref{conclusion}, we highlight the model limitations and anticipated model developments and present our conclusions.

\section{Chemistry and Thermodynamics}
\label{chem_therm}
In this work, we present a new model for ISM physics called ``{\small PRISM}'' that is implemented in the {\small RAMSES-RTZ}\footnote{{\small RAMSES-RTZ} is an extension of both the {\small RAMSES} \citep{Teyssier2002} and {\small RAMSES-RT} \citep{Rosdahl2013} codes.} code \citep{RTZ}. The model has many similar features to various 1D implementations of cooling and heating processes in the ISM \citep{Wolfire1995,Koyama2000,Wolfire2003,Bialy2019}. We describe the ingredients of our model below and compare with previous work in the literature to demonstrate that the model is behaving as expected. Our goal is to demonstrate that our 3D code can converge to the correct equilibrium solution, regardless of ISM conditions.

\subsection{Chemistry}
Our code can follow the formation and destruction of up to 115 chemical species. This includes the primordial species of H~{\small I}, H~{\small II}, He~{\small I}, He~{\small II}, He~{\small III}, and $e^-$ (6), all metallic ionization states of O, C, N, Fe, S, Si, Mg, and Ne (107), as well as H$_2$ and CO molecules (2). The species are fully coupled to on-the-fly radiation hydrodynamics via eight frequency bins (see Table~\ref{tab:engy_bins}) and to the thermodynamics of the gas via various cooling and heating processes described below. 

\begin{table}
    \centering
    \begin{tabular}{lllp{1.6in}}
    \hline
    Group Name & $E_{\rm low}$ & $E_{\rm high}$ & Function \\
    & (eV) & (eV) & \\
    \hline
    IR & 0.1 & 1.0 & Infrared radiation pressure \\
    Opt. & 1.0 & 5.6 & Direct radiation pressure \\
    
    FUV & 5.6 & 11.2 & Photoelectric heating, Mg~{\small I}, Si~{\small I}, S~{\small I}, Fe~{\small I} ionization \\
    
    LW & 11.2 & 13.6 & H$_2$ dissociation, C~{\small I} ionization \\
    
    EUV$_1$ & 13.6 & 15.2 & H~{\small I}, N~{\small I}, O~{\small I}, Mg~{\small II} ionization \\
    
    EUV$_2$ & 15.2 & 24.59 & H$_2$, C~{\small II}, Si~{\small II}, S~{\small II}, Fe~{\small II}, Ne~{\small I} ionization \\
    
    EUV$_3$ & 24.59 & 54.42 & He~{\small I}, O~{\small II}, C~{\small III}, N~{\small II}, N~{\small III},  Si~{\small III}, Si~{\small IV}, S~{\small III}, S~{\small IV}, Ne~{\small II}, Fe~{\small III} ionization \\
    
    EUV$_4$ & 54.42 & $\infty$ & He~{\small II}, O~{\small III}+, N~{\small IV}+, C~{\small IV}+, Mg~{\small III}+, S~{\small V}+, Si~{\small V}+, Fe~{\small IV}+, Ne~{\small III}+ ionization \\
    \hline
    \end{tabular}
    \caption{Properties of the radiation energy bins in the simulation. We list the photon group name, the lower and upper energies, and the main function. An ion is listed if its ionization potential falls within the energy range of the bin. Ions listed with a ``+'' symbol indicate that all higher ionization states of the element fall within the bin. }
    \label{tab:engy_bins}
\end{table}

As described in \cite{RTZ}, the code accounts for photoionization assuming cross sections from \cite{Verner1996}, collisional ionization using rates from \cite{Voronov1997}, radiative recombination and dielectronic recombination adopting rates from \cite{Badnell2006} and \cite{Badnell2003}, respectively, and charge exchange reactions following \cite{Kingdon1996}. A few reaction rates deviate from the default references provided here and these unique cases are well described in \cite{RTZ}. In addition to the above-mentioned processes, new for this work, we have also included recombination on dust grains \citep{Weingartner2001} as well as cosmic-ray ionization.

\subsubsection{H$_2$ formation and destruction}
H$_2$ formation and destruction follows the method presented in \cite{Katz2017} with minor modifications. We assume that H$_2$ forms via two channels: on dust grains, and through a primordial process that primarily\footnote{Note that we ignore three-body formation of H$_2$ as this process primarily occurs at densities much higher than we intend to probe with the code \citep[e.g.][]{Turk2011}.} involves H$^-$. The reaction rate for H$_2$ formation on dust ($R_{\rm d}$) is a combination of that presented in \cite{Gnedin2009} and \cite{Bialy2019}. We set
\begin{equation}
    R_{\rm d}=3.5\times10^{-17}f_{\rm dg}C\sqrt{\frac{T}{100\ {\rm K}}}\ \ \  {\rm [cm^3\ s^{-1}]},
\end{equation}
where $R_{\rm d}$ is empirically derived in \cite{Wolfire2008}, $f_{\rm dg}$ is the dust-to-gas mass ratio normalized to that of the Milky Way (see below), $C$ is a clumping factor that accounts for unresolved density structure below the grid scale\footnote{This parameter is often tuned to match an observed atomic-to-molecular transition. For this work, we will always assume $C=1$.}, and $T$ is the gas temperature. The primary modification in this work compared to \cite{Katz2017} are the explicit dependence of the reaction rate on temperature. Furthermore we adopt a $f_{\rm dg}$ that does not linearly scale with metallicity.

As we do not follow the non-equilibrium abundance of H$^-$, at every time step, we solve for the equilibrium abundance of H$^-$ by combining Equations 1, 2, 5, 13, 14, and 15 listed in Table~B1 of \cite{Glover2010}, with destruction rates of H$^-$ due to cosmic-rays and the local radiation field. The rates of H$^-$ destruction due to cosmic-rays and the local radiation field, which is new for this work, are scaled according to those used in \cite{Heays2017}.

Finally, we account for self-shielding of H$_2$ from the local radiation field following \cite{Gnedin2009}.

\subsubsection{CO formation and destruction}
Due to the inaccessibility of direct H$_2$ observations in many astrophysics environments, CO is often used to probe molecular gas \citep[e.g.][]{Bolatto2013}. CO is now readily observed at $z>5$ in a variety of sources \citep[e.g.][]{Riechers2019}. From a theoretical perspective, CO is expected to be an important coolant of the ISM in high-density gas \citep[e.g.][]{Neufeld1995} due to it being one of the most common diatomic molecules \citep{Solomon1972}. For these reasons, we include a model for CO formation and destruction in our chemical network. 

Our CO model is based on those presented in \cite{Nelson1997} and \cite{Glover2012}. The formation rate of CO follows:
\begin{equation}
    \frac{dn_{\rm CO}}{dt}=k_0n_{\rm C^+}n_{\rm H_2}\beta-\Gamma_{\rm CO}n_{\rm CO},
    \label{COform}
\end{equation}
where $n_{\rm CO}$ is the CO number density, $k_0=5\times10^{-16}$ \citep{Nelson1997}, $n_{\rm C^+}$ is the C$^+$ number density\footnote{In \cite{Nelson1997} they assume that all C is in the form of C$^+$. Because our chemical model follows individual C ionization states, we have relaxed this assumption and use the true C$^+$ number density in each cell.}, $n_{\rm H_2}$ is the molecular hydrogen number density, and $\Gamma_{\rm CO}=2.43\times10^{-10}G_0f_{\rm sh,CO}+4.62\times10^{-15}(\eta_{\rm cr}/10^{-16})$ is the destruction rate of CO which is the combination of destruction due to the UV radiation field and cosmic-ray ionization rate ($\eta_{\rm cr}$). We account for self-shielding ($f_{\rm sh,CO}$) by modulating the destruction due to the local radiation field ($G_0$) by interpolating tables from \cite{Lee1996} for CO, H$_2$ and dust, and measuring the column densities locally. Destruction coefficients are adopted from \cite{Heays2017}\footnote{Note that our CO destruction coefficient due to the local radiation field is slightly higher than that presented in \cite{Glover2012} due to a different choice of atomic data.}. 

CO formation is primarily governed by radiative association reactions that form CH$_x$ (e.g. CH, CH$_2$, etc.) molecules. Because we do not follow CH$_x$ molecules in our network, this physics\footnote{This refers to the initial radiative association reaction of ${\rm C^++H_2\rightarrow CH_2^+}$ which then converts to CH$_x$ via further reactions with H$_2$ and electrons.} is encapsulated in the coefficient $\beta$ in Equation~\ref{COform}.
\begin{equation}
    \beta = \frac{k_1x_{\rm O}}{k_1x_{\rm O}+\Gamma_{\rm CH_x}/n_{\rm H}},
\end{equation}
where $k_1=5\times10^{-10}$, $n_{\rm H}$ is the number density of hydrogen nuclei, $x_{\rm O}$ is the oxygen abundance fraction, and $\Gamma_{\rm CH_x}=1.41\times10^{-10}G_0+8.88\times10^{-15}(\eta_{\rm cr}/10^{-16})$, with destruction coefficients adopted from \cite{Heays2017}. Because we follow the fluxes of sub-ionizing radiation directly in the simulation, the radiative transfer algorithm in {\small RAMSES-RTZ} already accounts for the absorption of this radiation by dust. Hence we can neglect the explicit dependence of the destruction rates of CO and CH$_x$ on the dust optical depth.

\subsubsection{Metal abundances and production}
For models where we define an initial metallicity, we assume solar metal abundance patterns following \cite{Grevesse2010} and scale up and down depending on metallicity. For solar oxygen abundance, we use $12+\log_{10}({\rm O/H})=8.69$ and when referring to ``metallicity'', we specifically refer to that measured from the oxygen abundance. In simulations that include stellar particles or star formation, we also allow for the production of metals on-the-fly. Our model combines yields from Population~III and Population~II stars, via core-collapse supernova, hypernova (Pop.~III only), pair-instability supernova (Pop.~III only), type-Ia supernova, and AGB winds \citep{Portinari1998,Heger2002,Nomoto2006,Seit2013,Pignatari2016} following the model in \cite{Andersson2022}. Further description of the model and how we sample the stellar initial mass function is presented in \cite{Katz2022-popiii}.

\subsubsection{Dust model}
While we do not currently follow the creation and destruction of dust on-the-fly, we employ an effective model for dust that is used for heating and cooling, attenuating the radiation field, radiation pressure, and metal depletion. Following earlier work \citep{Kimm2018}, we assume that the dust-to-gas mass ratio scales with metallicity based on the empirical results of \cite{RR2014}. More specifically, we adopt their $X_{\rm CO,Z}$ model fitted with a broken power-law. To be consistent with \cite{RR2014}, the dust composition is assumed to be that of the BARE-GR-S model of \cite{Zubko2004}.

\begin{table}
    \centering
    \begin{tabular}{l|lllllllll}
    \hline
         Metal & Fe & O & N & Mg & Ne & Si & Ca & C & S \\
         \hline
         $f_{\rm gas}$ & 0.01 & 0.73 & 0.6 & 0.16 & 1 & 0.1 & 0.003 & 0.5 & 1 \\
    \hline
    \end{tabular}
    \caption{Gas-phase fractions of each metal at solar metallicity. We do not allow the metals to be depleted below these fractions. Values of one indicate that the metal is always in the gas phase and hence not included in the dust model.}
    \label{tab:dep}
\end{table}

The combination of a dust-to-gas mass ratio and dust composition sets the dust cross section as a function of wavelength which is used for attenuating the radiation field and for radiation pressure, the mass of metals locked up in dust (depletion), and the PAH abundance, which is needed for recombination rates on dust as well as dust heating and cooling. Since the dust-to-gas mass ratio is defined based on the oxygen abundance and the dust composition is fixed, differential metal production in our simulations can sometimes lead to unphysical depletion values (i.e. more of a particular metal is expected to be depleted than is available in the gas cell). To avoid this issue, we place an upper limit on the fraction that each metal can be depleted which reflects their gas-phase fractions at solar metallicity. These minimum allowed gas-phase percentages can be found in Table~\ref{tab:dep} and are a combination of the depletion values from \cite{Dopita2000} and the fraction of a metal expected to be depleted at solar metallicity by combining the \cite{Grevesse2010} abundances with the BARE-GR-S dust composition. Note that our method does not explicitly conserve the total mass of each element due to the fixed dust composition and this will be remedied in future work where we explicitly model dust production and destruction on-the-fly (Rodriguez~Montero~et~al.~{\it in prep}).

\subsection{Heating \& Cooling}
The total rate of change of energy density ($E$) in each gas element is defined as
\begin{equation}
\dot{E}=\sum_{k}^{{\rm processes}}\sum_{i}^{{\rm species}} n_i\Gamma_k-\sum_{k}^{{\rm processes}}\sum_{i}^{{\rm species}}\sum_{j}^{{\rm species}} n_{i}n_{j}\Lambda_k,
\end{equation}
for different combinations of species $i$ and $j$, where the total heating rate is the sum of heating ($\Gamma$) due to cosmic rays, photoheating, photoelectric heating, and H$_2$ formation and dissociation heating. The net cooling rate is the sum of cooling ($\Lambda$) from primordial species (and their ions) such as H, He, and H$_2$, metal line cooling such as fine-structure transitions of O~{\small I} and C~{\small II}, CO molecular cooling, and dust processes such as recombination or grain-gas collisional cooling. Below we discuss the details of how each of these quantities is implemented in the {\small PRISM} model.

\subsubsection{Cosmic-ray heating}
Due to their long mean-free paths, cosmic-ray (and X-ray) ionization is the primary mechanism that sets the electron fraction at high densities. Cosmic-rays are also expected to dominate the heating rates at low densities, where photoelectric heating becomes less efficient. Our implementation of cosmic-ray interactions with primordial species follows \cite{Bialy2019} (see their Equations 37-39). We consider both primary and secondary ionizations and the amount of heating per ionization event is adopted from \cite{Dalgarno1972,Draine2011}. Free electrons can also directly interact with cosmic-ray protons. We adopt a heating rate due to this interaction following \cite{Goldsmith1969,Bialy2019}.

In addition to primordial species, we consider cosmic-ray heating due to interactions with metals. Primary cosmic-ray ionization rates for metals are based on \cite{Richings2014}, following the methods of \cite{Lotz1967,Silk1970,Langer1978}. For these ionizations, we adopt the same heating rates as for primordial species. Due to the low number densities of metals, this mechanism is generally unimportant. However, interactions between cosmic-rays and H$_2$ can induce a UV field \citep{Gredel1987,Gredel1989} that leads to orders of magnitude more photodestruction of molecules such as CO or photoionization of metals with low ionization potentials compared to the primary cosmic-ray ionziation rate. Photoionization and destruction rates from this induced UV flux are adopted from \cite{Heays2017} and the corresponding photoheating rates are calculated by differencing the mean photon energy of the induced spectrum above the relevant ionization potential. 

For our fiducial model, we adopt a primary cosmic-ray ionization rate of $10^{-16}\ {\rm s^{-1}\ H^{-1}}$ \citep[e.g.][]{Indriolo2012}, which is uniform in every cell of the simulation.

\subsubsection{Photoheating}
When photons ionize an atom or a molecule, the excess energy of the photon above the ionization potential is converted into kinetic energy that heats the gas. Since our radiation field is discretized, we follow Equation A16 in \cite{Rosdahl2013}. In principle, the average photon energy in each bin should only consider local sources around the gas cell of interest. However, for computational efficiency, we assume that the average photon energy in each bin is a luminosity-weighted sum of all sources in the computational volume. 

\subsubsection{Photoelectric heating}
Photoelectric heating is one of the dominant ISM heating processes at solar metallicity. Our implementation follows that presented in \citealt{Kimm2017} (see Equations 25 and 26), which is based on \cite{Bakes1994,Wolfire1995,Wolfire2003}. Because we assume a different dust model compared to \cite{Kimm2017}, we have modified our original implementation in two ways. First, we scale the heating rate with an empirically calibrated dust-to-gas mass ratio as a function of metallicity \citep[see above,][]{RR2014}. Second, the normalization of the heating rate is dependent on the number of C atoms locked up in polycyclic aromatic hydrocarbons (PAHs). \cite{Wolfire2003} assume this to be $22\times10^{-6}$ per H atom \citep{Tielens1999}. However, for our adopted BARE-GR-S model from \cite{Zubko2004}, the number of C atoms locked up in PAHs is $33\times10^{-6}$ per H atom. Thus we increase the heating rate compared to \cite{Wolfire2003} by a factor of 1.5.

\subsubsection{H$_2$ heating}
We consider three mechanisms by which H$_2$ heats the gas: formation, dissociation, and pumping. When H$_2$ forms, a fraction of the binding energy is released into the gas in the form of heat. Following \cite{Sternberg1989,Rollig2006}, we assume that $1/3$ of the 4.5~eV binding energy is released as heat.

When H$_2$ is dissociated due to Lyman-Werner band photons, we assume that 0.4~eV is deposited as heat into the gas \citep{Black1977}. Only $\sim15\%$ of Lyman-Werner band photons are expected to dissociate H$_2$, while the other $\sim85\%$ simply excite the molecule \citep[e.g.][]{Draine1996}. If the molecule is collisionally de-excited, the Lyman-Werner band photon is converted into heat. Otherwise, it is radiated away. Our model for heating due to H$_2$ pumping follows Equations $44-48$ in \cite{Fervent} that combines the UV pumping rate of H$_2$ from \cite{Draine1996} with the energy released per UV pumping event from \cite{Burton1990}.

We also consider photoionization of H$_2$ by photons with $E>15.2$eV, which we assume always leads to dissociation via recombinative-dissociation. As with all photoionization, the excess energy of the photon above the ionization potential of the molecule is converted into heat. 

\subsubsection{Primordial atomic cooling}
At $T\gtrsim8000$~K, transitions in primordial atomic species (i.e. H and He), in particular Ly$\alpha$, often dominate the total cooling rate. Our implementation is the same as presented in Appendix E3 of \cite{Rosdahl2013}. This includes contributions from collisional ionization cooling \citep{Cen1992}, radiative recombination cooling \citep{Hui1997}, collisional excitation cooling \citep{Cen1992}, Bremsstrahlung cooling \citep{Osterbrock2006}, Compton cooling/heating \citep{Haiman1996}, and dielectronic recombination cooling \citep{Black1981}.

\subsubsection{H$_2$ cooling}
At low metallicity ($\lesssim10^{-5}Z_{\odot}$), H$_2$ cooling is the dominant process that allows the gas to cool below $\sim10^4$~K \citep[e.g.][]{Saslaw1967,Yoneyama1972,Lepp1984}. For H$_2$ cooling, we adopt the fits to tabulated data presented in \citealt{Galli1998} (see their Appendix A). The cooling function follows the form presented in \cite{Hollenbach1979}, accounting for the change in cooling above and below the critical density. 

\subsubsection{Metal-line cooling}
In metal-enriched gas, metal-line emission is often the dominant coolant of the ISM. Our method is identical to that presented in \cite{RTZ} and we briefly describe it here. At $T<10^4$~K, metal-line cooling is dominated by fine-structure transitions. We analytically calculate the level populations for O~{\small I}, O~{\small III}, C~{\small I}, C~{\small II}, N~{\small II}, Fe~{\small I}, Fe~{\small II}, Si~{\small I}, Si~{\small II}, S~{\small I}, and Ne~{\small II}, assuming either a two-level or three-level system where relevant. Furthermore, we include stimulated emission coefficients in our level-population calculation to account for the CMB radiation field. We have updated all collisional partner data for the following collision partners: H, H$^+$, ortho and para H$_2$, $e^-$, He, He$^+$, and He$^{++}$ to be consistent with {\small CLOUDY} \citep{Ferland2017}. At $T>10^4$~K, the dominant collisional partner for metal-line emission is electrons. For this reason, we adopt tabulated values for ion-by-ion cooling rates from \cite{Oppenheimer2013}. In order to smoothly interpolate between the two temperature regimes, we have adopted the method of \cite{Bovino2016}. The metal cooling rate is defined as
\begin{equation}
    \Lambda_Z=f_1(T)\Lambda_Z^{\rm tabulated}+f_2(T)\Lambda_Z^{\rm fine-structure},
\end{equation}
where
\begin{equation}
\begin{split}
f_1(T)=0.5[\tanh(c_{\rm sm}(T-10^4))+1]\\
f_2(T)=0.5[\tanh(c_{\rm sm}(10^4-T))+1],
\end{split}
\end{equation}
which smoothly interpolates between the fine-structure and tabulated cooling at $T\sim10^4$~K. To limit overlap, we have increased the $c_{\rm sm}$ parameter from $10^{-3}$ as was used in \cite{Bovino2016} to $5\times10^{-3}$.

Our model allows for a reduced chemical network where not all metal ionization states are self-consistently tracked. More specifically, for computational efficiency, the most highly ionized state followed by our code is often less than the total number of possible ionization states. In this case, the most highly ionized state that we follow represents the total amount of gas in that ionization state or higher. For example, if we only track O~{\small I}-O~{\small VI} in the code, O~{\small VI} really represents the ionization states of O~{\small VI}-O~{\small IX}. To calculate cooling in the situation where not all ionization states are tracked, we assume collisional ionization equilibrium (CIE) for all untracked higher ionized states and use the ion-by-ion cooling tables of \cite{Oppenheimer2013}. In our example, the distribution of O in O~{\small VI}-O~{\small IX} will be calculated in CIE. A similar method was adopted by \cite{Gray2015}.

\subsubsection{CO cooling}
Rotational and vibrational excitations of CO can lead to radiative cooling. We model both processes following \citealt{Koyama2000} (see their Equations A10 and A11). Rotational cooling in the optically thin limit is derived from \cite{Mckee1982}, while vibrational cooling rates are adopted from \cite{Hollenbach1989}.

\subsubsection{Dust recombination cooling}
Charged particles can stick to the surfaces of PAHs which leads to gas cooling \citep[e.g.][]{Draine1987,Bakes1994}. We have implemented this cooling process following \citealt{Wolfire2003} (see their Equation 21). Similar to our discussion of photoelectric heating, the assumed PAH abundance in our dust model is 50\% higher than that of \cite{Wolfire2003}, so we proportionally scale the cooling rate by a factor of 1.5.

\subsubsection{Grain-gas collisional cooling}
When the gas temperature is greater than the dust temperature, collisional interactions can lead to gas cooling \citep{Hollenbach1989}. Different sized grains are expected to exhibit different temperatures; however, modelling a self-consistent grain size distribution and its coupling to the radiative transfer and gas is beyond the scope of this work. For simplicity, we follow \cite{Draine2011,Bialy2019} and assume that the dust temperature $T_{\rm dust}=17.9\ {\rm K}\ G_0^{1/6}$, where $G_0$ is the flux of the FUV radiation field normalized to a value of $1.6\times10^{-3}\ {\rm erg\ s^{-1}\ cm^{-2}}$ \citep{Habing1968}. We adopt a fixed grain size of 100~\AA. The cooling (or heating) rate is then calculated following Equation 2.15 in \cite{Hollenbach1989}, with a modification to account for the changing dust-to-gas-mass ratio with metallicity.

\subsection{Model Validation}
To validate our model, we initialize a grid of 512 gas cells that smoothly interpolate density between $10^{-3}-10^{5}$~cm$^{-3}$ in log-space. All cells have an initial temperature of $10^4$~K and are in an entirely neutral and atomic state. For our fiducial model, we set $G_0=1$, $C=1$, and $\eta_{\rm cr}=10^{-16}\ {\rm s^{-1}\ H^{-1}}$. We run the simulation six times, varying the metallicity in the range $10^{-5}Z_{\odot}-Z_{\odot}$ in steps of 1~dex. The gas density is fixed throughout the calculation and all simulations are evolved until equilibrium is reached.

\begin{figure}
    \centering
    \includegraphics{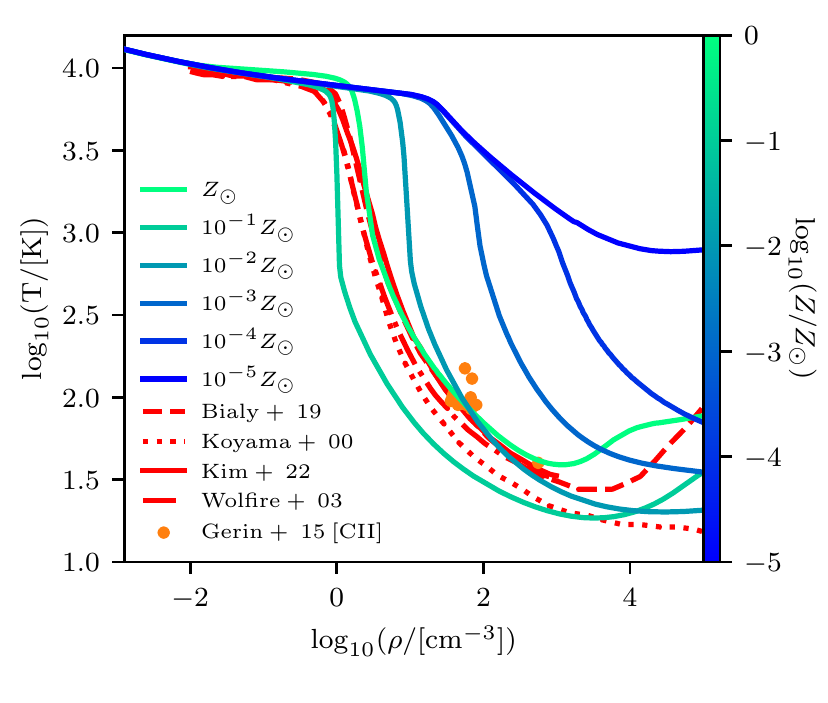}
    \includegraphics{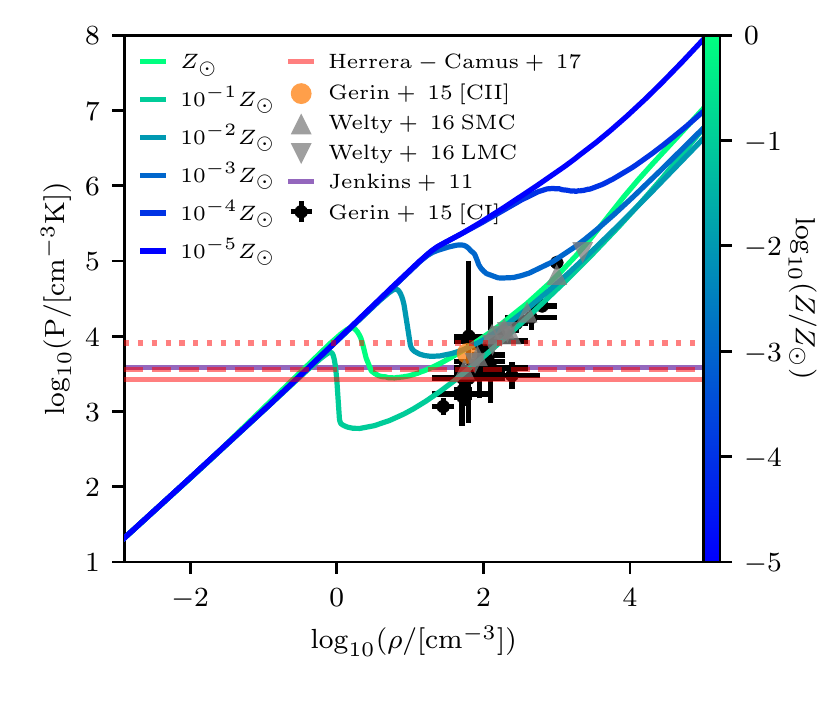}
    \caption{(Top) Equilibrium temperature-density relations as a function of metallicity. The dashed, dotted, dashed-dotted, and solid red lines show the results from \protect\cite{Bialy2019}, \protect\cite{Koyama2000}, \protect\cite{Wolfire2003}, and \protect\cite{JKim2022} respectively, at solar metallicity. Orange points represent [C~{\small II}] observational constraints in the galactic plane from \protect\cite{Gerin2015}. (Bottom) Equilibrium pressure-density relations as a function of metallicity. We compare our results to the galactic plane pressure measurements from \protect\cite{Gerin2015} using [C~{\small I}] (black) or [C~{\small II}] (orange) emission, from \protect\citep{Jenkins2011} (purple), and those from \protect\cite{Herrera2017} for different CNM fractions of 0.7, 0.5, and 0.3 as red dashed, solid, and dotted lines, respectively. LMC and SMC constraints from \protect\cite{Welty2016} are shown as grey triangles.}
    \label{fig:trho}
\end{figure}

In Figure~\ref{fig:trho} we show the equilibrium temperature (top) and pressure (bottom) as a function of gas density for the simulations with different metallicities (different colours). We note a few important features. First, at $Z\gtrsim10^{-3}Z_{\odot}$, the pressure curve exhibits a canonical `S' shape, as first described by \cite{Field1969}, where gas at different temperatures (and densities) can co-exist at the same pressure. This indicates that the ISM is multiphase. Similar to others \citep[e.g.][]{Norman1997,Bialy2019}, we find that this behaviour weakens and the gas evolves to higher pressures as metallicity goes to zero. Below $10^{-5}Z_{\odot}$, the gas is no longer multiphase. The reason for this is easily observed in the top panel of Figure~\ref{fig:trho} where at low-metallicity, the strength of the thermal instability is significantly suppressed.

\begin{figure}
    \centering
    \includegraphics{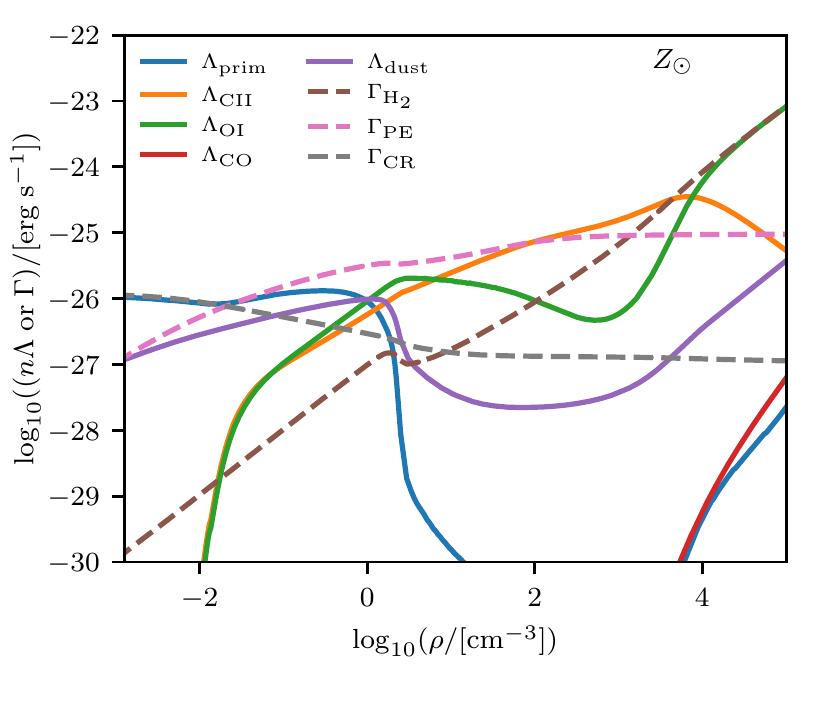}
    \includegraphics{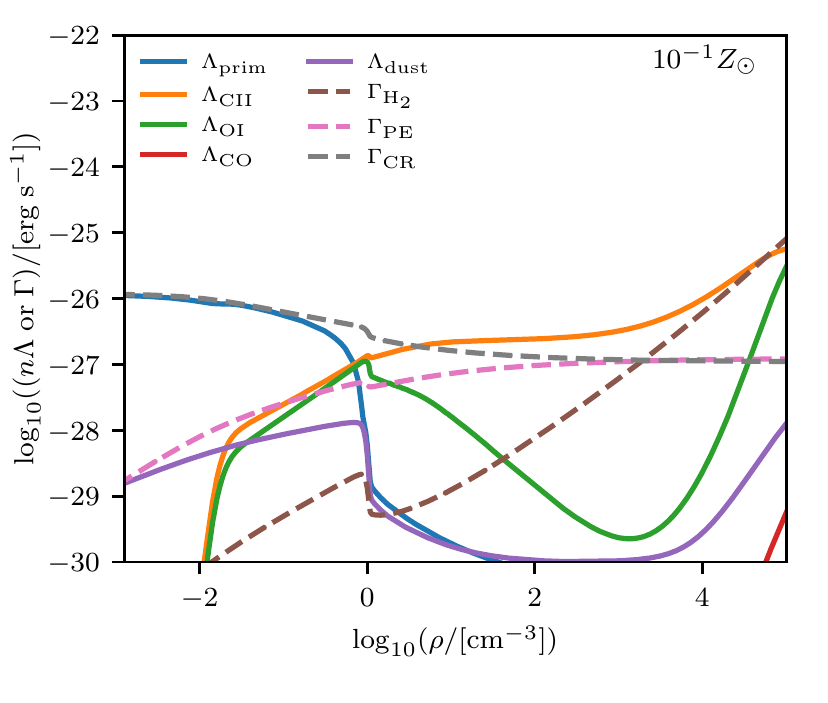}
    \includegraphics{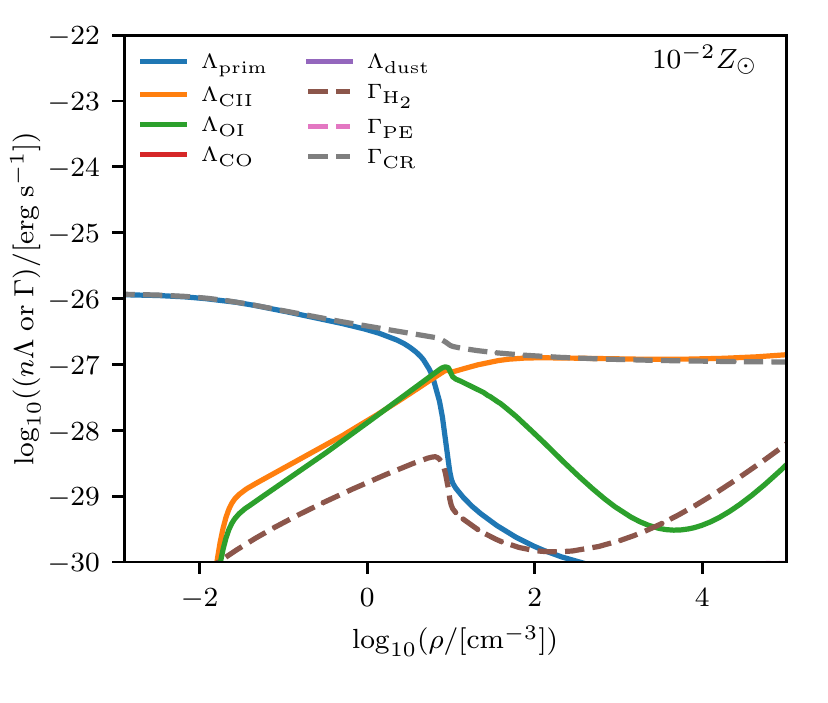}
    \caption{Equilibrium heating (dashed lines) and cooling (solid) rates for various processes as a function of gas density for three different metallicities as indicated on each panel. Different colours represent different processes including primordial cooling from H, He, and H$_2$ (blue), C~{\small II} fine-structure cooling (orange), O~{\small II} fine-structure cooling (green), CO molecular cooling (red), dust recombination and dust-gas collisional cooling (purple), H$_2$ formation and destruction heating (brown), photoelectric heating (pink), and cosmic-ray heating (gray).}
    \label{fig:heatcool}
\end{figure}

For comparison, we show the solar-metallicity temperature-density curves of \cite{Koyama2000}, \cite{Wolfire2003}, \cite{Bialy2019}, and \cite{JKim2022} as the dotted, dashed, and solid red lines, respectively in Figure~\ref{fig:trho}. Our model is most similar to that of \cite{Bialy2019} and indeed we find the two curves have the same general behaviour. Thermal instability is triggered at $\rho\gtrsim1\ {\rm cm^{-3}}$, primarily due to O~{\small I} and C~{\small II} cooling. This can be seen in Figure~\ref{fig:heatcool} where we show the heating and cooling rates for different processes as a function of density for three different metallicities. At high densities at solar metallicity, we see an increase in temperature (see Figure~\ref{fig:trho}). This is driven by H$_2$ heating (see the top panel of Figure~\ref{fig:heatcool}). Mild discrepancies between our curves and those of \cite{Bialy2019} are likely due to differences in metal abundances\footnote{\cite{Bialy2019} assume a solar O and C abundance of $5.4\times10^{-4}$ and $3.0\times10^{-4}$, respectively, while \cite{Koyama2000} assume $4.6\times10^{-4}$ and $3.0\times10^{-4}$. The fiducial abundances we adopt from \cite{Grevesse2010} are $4.9\times10^{-4}$ and $2.69\times10^{-4}$ for O and C, respectively.}, dust depletions, and reaction and heating/cooling rates. 

Our solar metallicity model is consistent with observational constraints of the pressure in the galactic plane. This includes data from \cite{Herrera2017} (assuming a CNM fraction of 0.3) and from \cite{Gerin2015}, both their [C~{\small I}] and [C~{\small II}] observations. The clear amount of scatter in the observational data suggests that there is no single equilibrium curve that is fully representative. Modulating the abundances/depletion factors, cosmic ray ionization/heating rates, and photoelectric heating rates within the uncertainties for the galaxy can bring the pressure curve into agreement with any of this observational data (which itself is subject to modelling uncertainties).

Once the metallicity is decreased by a factor of ten, thermal instability occurs at lower densities compared to solar. This is because at this metallicity, the dust-to-gas mass ratio is lowered by more than a factor of ten, which both reduces the amount of depletion of metals, and reduces the impact of photoelectric heating (as described in \citealt{Bialy2019}). The middle panel of Figure~\ref{fig:heatcool} shows the strength of the primary heating and cooling processes at $10^{-1}Z_{\odot}$ and for our model, cosmic-ray heating always dominates over photoelectric heating at this metallicity. The pressure curve for $10^{-1}Z_{\odot}$ is predicted to be lower than the solar metallicity case (see Figure~\ref{fig:trho}). Interestingly, this prediction is in agreement with the thermal pressure constraints in the LMC and SMC from \cite{Welty2016}, shown as grey triangles in Figure~\ref{fig:trho}

Moving to even lower metallicities, the thermal instability occurs at higher densities, until cooling is completely dominated by H$_2$ (between $10^{-4}Z_{\odot}-10^{-5}Z_{\odot}$). There is a stark contrast between high- and low-metallicity in that at high metallicities, H$_2$ primarily acts to heat the gas while at low-metallicities, it is responsible for cooling \citep[see e.g.][]{Omukai2000,Omukai2005}. By this point, photoelectric heating is negligible and the only heating terms that matter in this test are those from cosmic-rays and H$_2$.

\begin{figure}
    \centering
    \includegraphics{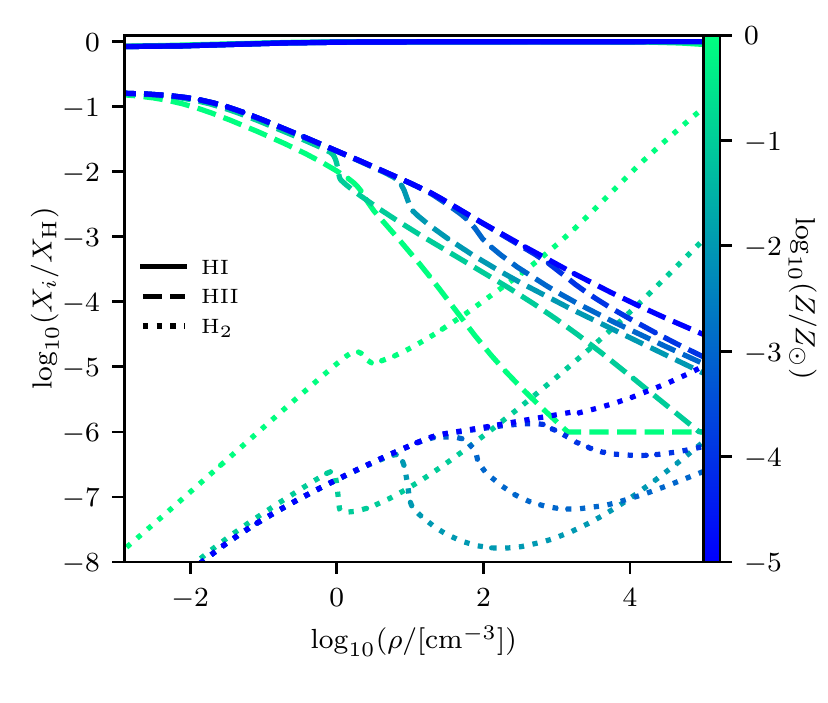}
    \includegraphics{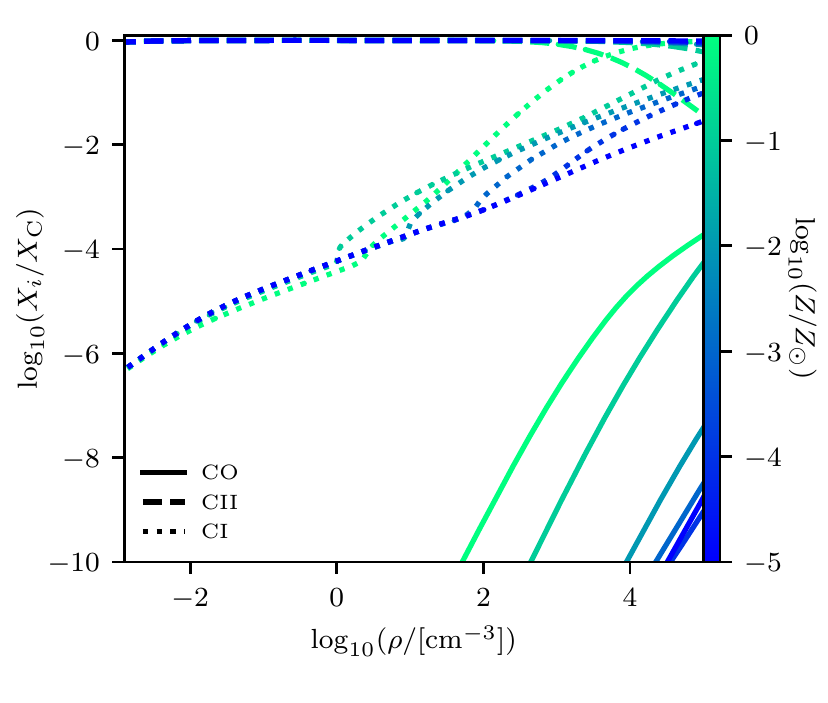}
    \caption{(Top) Equilibrium distribution of hydrogen in the form of H~{\small I} (solid), H~{\small II} (dashed), and H$_2$ (dotted) as a function of metallicity. (Bottom) Equilibrium distribution of carbon in the form of CO (solid), C~{\small II} (dashed), and C~{\small I} (dotted) as a function of metallicity.}
    \label{fig:mol}
\end{figure}

In Figure~\ref{fig:mol}, we show the fraction of H that is in the form of H~{\small I}, H~{\small II}, and H$_2$ (top panel) as well as the fraction of C that is in the form of C~{\small I}, C~{\small II}, and CO (bottom panel). Due to the lack of ionizing radiation, the hydrogen ionization fractions are set by collisional ionization and cosmic-ray ionization. For this reason, most of the H is neutral. At high enough densities and metallicities, recombination of H~{\small II} on dust grains becomes efficient and the solar metallicity curve for H~{\small II} begins to deviate from the other metallicities at $\rho=1\ {\rm cm^{-3}}$. This curve even reaches the H~{\small II}-floor of $10^{-6}$ which is set for solver stability (but has no impact on the thermodynamics of the cloud). 

The H$_2$ fraction exhibits strong evolution with both density and metallicity. At high metallicities, H$_2$ formation is primarily driven by the dust-formation channel while at low metallicities, formation switches to the H$^{-}$ channel. At densities of $10^5\ {\rm cm^{-3}}$, the H$_2$ mass fraction saturates at $\sim10\%$ at solar metallicity. Note that these simulations apply a uniform radiation field with no self-shielding. Our 3D code accounts for self-shielding and thus this test is not fully representative of what we might expect in a 3D simulation as column densities cannot be measured in these one-zone tests.

As described above, the addition of CO is one of the major updates for this work compared to our earlier chemical networks. The bottom panel of Figure~\ref{fig:mol} shows that even at high densities, CO remains subdominant compared to other forms of C. Apart from solar metallicity, the vast majority of C is in the form of C~{\small II}. This is because the ionization potential of C is $<13.6$~eV and it can thus be ionized by the interstellar radiation field. However, at solar metallicity, recombination on dust can become important and at $\rho\gtrsim10^3\ {\rm cm^{-3}}$, we find that C~{\small I} dominates over C~{\small II}. Since our CO model requires C~{\small II} for formation, the rate of increase of CO slows down above these densities at solar metallicity. This is likely one of the reasons that we produce slightly less CO at high densities compared to \cite{Bialy2019}. Nevertheless, when self-shielding is included in 3D simulations, we expect significantly more CO to form \citep{Glover2010,Glover2012}.

The results shown in Figures~\ref{fig:trho}, \ref{fig:heatcool}, and \ref{fig:mol} are comparable to the 1D models presented in \cite{Wolfire1995,Koyama2000,Wolfire2003,Bialy2019,JKim2022}. The novel aspect of our model is that it is fully coupled to a 3D radiative hydrodynamics infrastructure that can be used to simulate realistic galaxies. The results above should provide confidence in our implemented ISM model and below, we exploit the 3D nature of our code to demonstrate the utility of our new framework.

\section{Numerical Simulations}
\label{sims}
As a first application of the {\small PRISM} model, we run high-resolution simulations of isolated dwarf galaxies to determine how the intuition gained from our 1D models translates to a 3D setting with full galaxy formation physics, where cooling, heating, and ionisation states are computed in non-equilibrium with on-the-fly radiative transfer, self-shielding, stellar feedback, metal production, and dynamics.

\begin{figure*}
    \centering
    \includegraphics{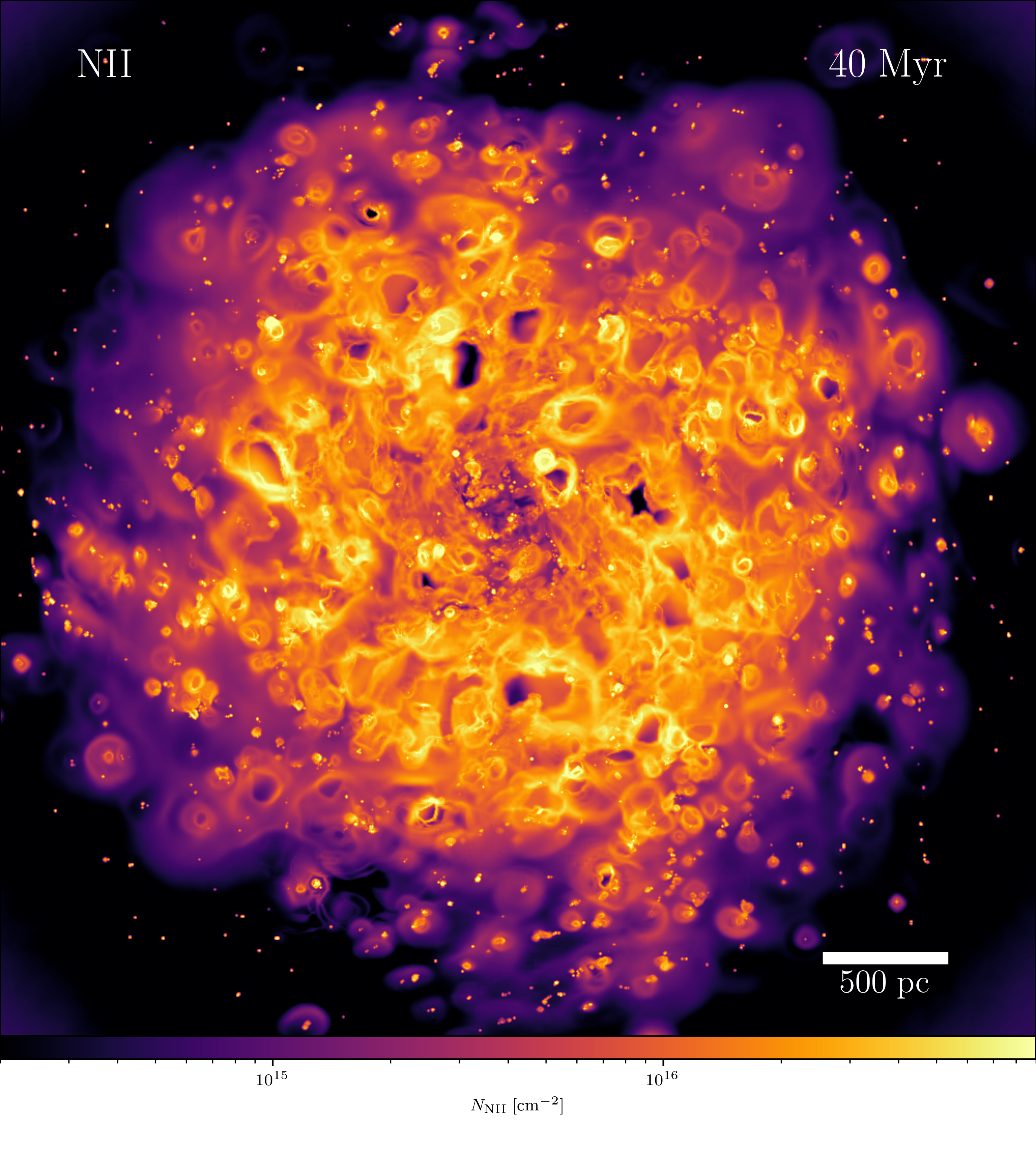}
    \caption{Column density map of N~{\small II} in the central regions of G9 at 40~Myr. At this time, the star formation rate is 1.78~${\rm M_{\odot}\ yr^{-1}}$. The ionization potential of N~{\small I} is 14.5~eV so the map is tracing the H~{\small II} regions around young stars as well as gas heated and enriched by supernova feedback. The diverse network of structure highlights the expected complexity and interactions between star-forming regions in real galaxies.}
    \label{fig:ions_11}
\end{figure*}

\subsection{Methods}
The simulations presented in this work are similar to those described in \cite{RTZ}. Here, we provide a brief overview of the included physics and highlight the differences compared to our previous work.

The simulations follow the evolution of two isolated, rotating disk galaxies, named G8 and G9. The initial conditions for these galaxies are identical to those presented in \citealt{Rosdahl2015} (see their Table~1). G8 and G9 have halo virial masses of $10^{10}\ {\rm M_{\odot}}$ and $10^{11}\ {\rm M_{\odot}}$, and circular velocities of $30\ {\rm km\ s^{-1}}$ and $65\ {\rm km\ s^{-1}}$, respectively. The disk gas masses are $3.5\times10^{8}\ {\rm M_{\odot}}$ and $3.5\times10^{9}\ {\rm M_{\odot}}$, both of which are embedded in a background gas density of $10^{-6}\ {\rm cm^{-3}}$. The initial central metallicity of G8 and G9 are $10^{-1}Z_{\odot}$ and $0.5Z_{\odot}$ and we assume that the metallicity decreases with radius and height above the disk\footnote{In practice, we set $Z=Z_{\rm ini}\times10^{0.5-r/r_{\rm cut}}$, where $r_{\rm cut}=5$~kpc for G8 and 11~kpc for G9.}. Initial chemical abundance ratios are scaled to the solar metallicity composition of \cite{Grevesse2010}. To reduce computational expense, we employ a reduced chemical network, following eight ionization states of O, seven of N, six of C, Si, Mg, Fe, S, and Ne, as well as all ionization states of H, and He, in addition to the H$_2$ and CO molecules. The initial composition of the disk is assumed to be atomic and neutral, while outside the disk, all gas is assumed to be fully ionized.

The major difference between the simulations in \cite{RTZ} and those presented here are that we adopt the {\small PRISM} model for all ISM cooling and heating processes and use the updated chemistry as described in Section~\ref{chem_therm}\footnote{Note that the simulations presented here use the old collisional rates from \cite{RTZ} for metal fine-structure cooling compared to the ones presented above and do not include the factor 1.5 boost for photoelectric heating and dust cooling due to the enhanced PAH abundance. The impact of these differences are discussed in Appendix~\ref{app:changes}.}. The simulations do not self-consistently follow cosmic-ray production or propagation so we employ a fixed cosmic-ray background ionization rate of $\eta_{\rm cr}=10^{-16}\ {\rm s^{-1}\ H^{-1}}$. Unlike our 1D calculations, radiation hydrodynamics is self-consistently followed in eight energy bins (see Table~\ref{tab:engy_bins}). A reduced speed of light approximation is used ($c_{\rm sim}=c/100$) to reduce computational expense. We consider two sources of radiation. The first is a UV background from \cite{Haardt2012}, which is spatially uniform throughout the computational volume. We apply a self-shielding approximation such that the intensity of the UV background is exponentially suppressed at $\rho>10^{-2}\ {\rm H\ cm^{-3}}$ in H-ionizing radiation energy bins. The second source of radiation is star particles. Stars can form in the simulation following a thermo-turbulent star formation criteria \citep{Federrath2012,Kimm2017}. Full details on the star formation recipe are presented in \cite{Rosdahl2018}. Star particle masses are integer multiples of 1830~M$_{\odot}$. Radiation is emitted from star particles based on their mass, age, and metallicity following a {\small BPASS~v2.2.1} SED \citep{Stanway2018}. Feedback from stars is modelled for SNII, SNIa, and stellar winds \citep{Kimm2015,Agertz2021}. 

The second important difference between the simulations presented here compared to those in \cite{RTZ} is spatial resolution. Motivated by the work of \cite{Kim2017} who demonstrated that many ISM properties converge in their model if the size of a gas cell is $\leq8$~pc, we allow the AMR grid in our simulations to refine up to a maximum resolution of 4.5~pc. Refinement in the simulation is triggered when the cell hosts eight times its initial mass in gas or to resolve the local Jeans length by four cells. This resolution is substantially improved compared to the 18~pc resolution used in \cite{RTZ}. We exemplify this in Figure~\ref{fig:ions_11} where we show an N~{\small II} column density of the central regions of G9 after 40~Myr, which is during the initial starburst phase of the galaxy. The N~{\small II} is highly structured and follows young star-forming regions.

In what follows, we will focus on understanding how the results from the simulations differ from the 1D models. 

\begin{figure}
    \centering
    \includegraphics{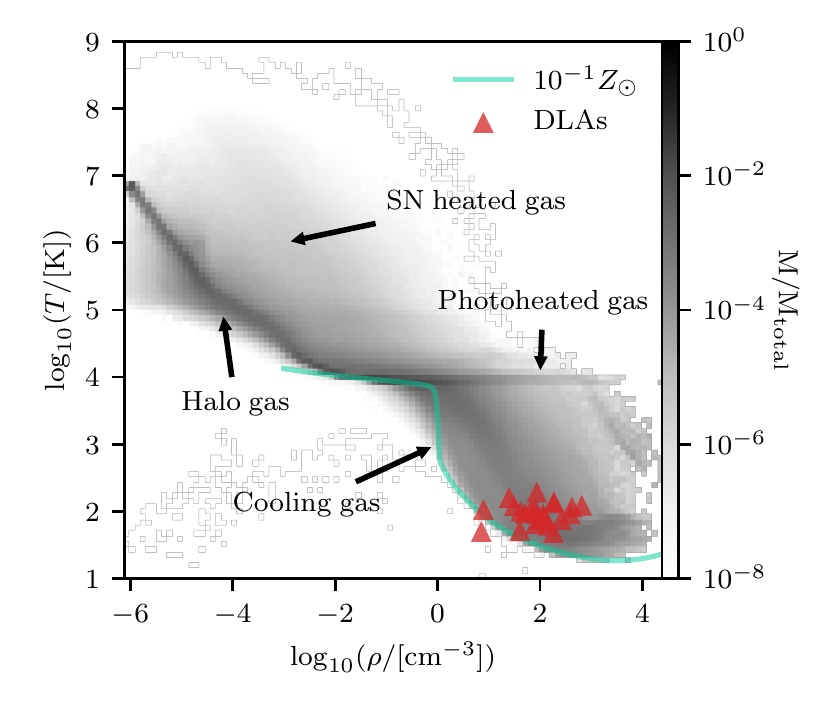}
    \includegraphics{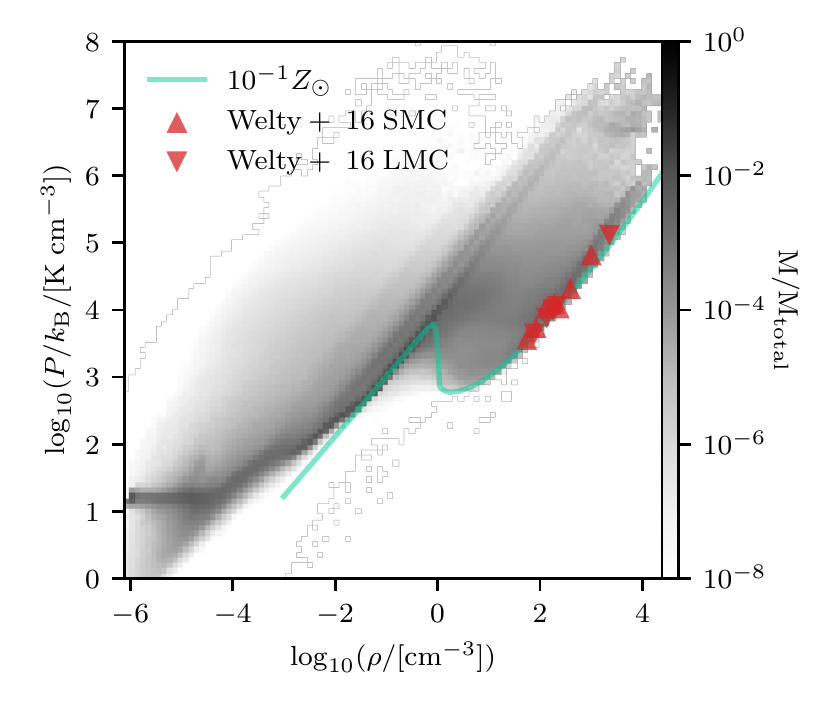}
    \caption{(Top) 2D mass-weighted histogram of temperature versus density for a stack of G8 snapshots from 150-400~Myr. We show density-temperature estimates for a sample of low- and high-redshift DLAs from \protect\cite{Balashev2019,Klimenko2020} that are representative of low-metallicity systems. (Bottom) 2D mass-weighted histogram of pressure versus density for a stack of G8 snapshots from $150-400$~Myr. For comparison, we show pressure measurements for LMC and SMC sight-lines from \protect\cite{Welty2016} as red triangles. In both plots, the cyan line shows the location of our 1D equilibrium model at $10^{-1}Z_{\odot}$, the initial metallicity of the galaxy.}
    \label{fig:sim_trho}
\end{figure}

\begin{figure}
    \centering
    \includegraphics{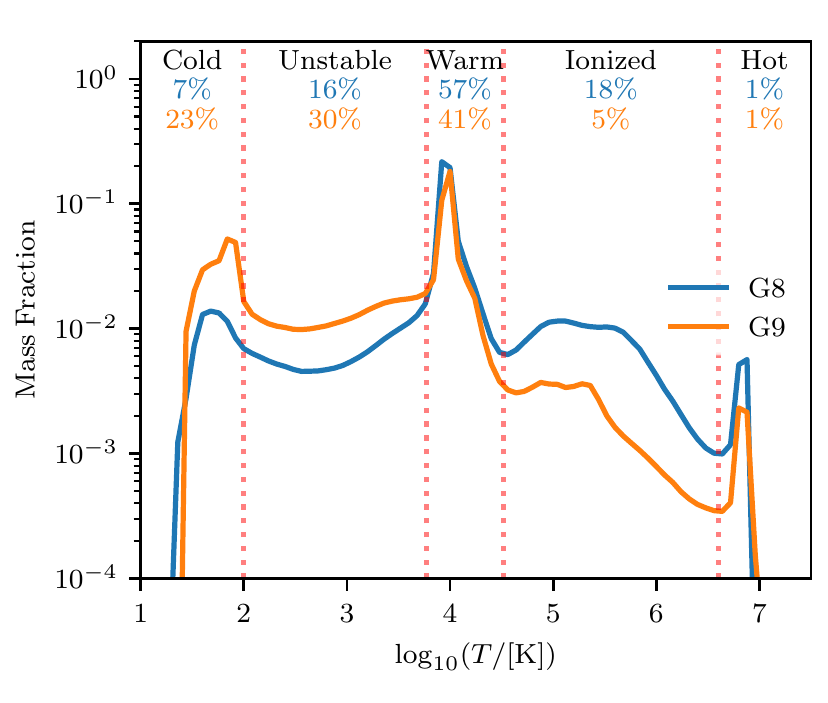}
    \caption{Mass fractions of gas as a function of temperature for a stack of G8 snapshots from 150-400~Myr (blue) and G9 snapshots from 150-200~Myr (orange). We differentiate phases of the ISM as cold ($T<100$~K), unstable ($100~{\rm K}\leq T<5,888~{\rm K}$), warm ($5,888~{\rm K}\leq T<3.3\times10^4~{\rm K}$), ionized ($3.3\times10^4~{\rm K}\leq T<4\times10^6~{\rm K}$), and hot ($T\geq4\times10^6~{\rm K}$) based on natural breaks in the distribution function. The mass fraction of total gas in each phase is listed.}
    \label{fig:sim_mf}
\end{figure}

\subsection{The density, pressure, and temperature distribution in the ISM}
Due to the dynamic nature of the simulation, we expect the density, pressure, and temperature of the ISM to strongly deviate from the 1D models. In Figure~\ref{fig:sim_trho} we show mass-weighted 2D histograms\footnote{We only consider gas with an oxygen mass fraction $>10^{-8}$ which removes much of the pristine, artificial CGM in place in the initial conditions.} of temperature versus density (top) and pressure versus density (bottom) for a stack\footnote{We calculate the total mass of within each 2D bin across all snapshots and compute the PDF.} of G8 snapshots from $150-400$~Myr\footnote{We begin our analysis after 150~Myr to limit the impact of the disk settling from the initial conditions.}. At high densities and low temperatures, rather than a single track, we find that the gas distribution has bifurcated into two equilibrium trajectories (the same behaviour is seen for G9) and these are annotated as ``cooling gas'' and ``photoheated gas''. The first extends towards low temperatures which represents the same equilibrium cooling curve that we find in the 1D models. There is considerably more scatter due to the fact that the metallicity varies between gas cells as does the sub-ionizing radiation field and the electron density. The second is gas that maintains a temperature of $\sim10^4$~K even at high densities. For this second trajectory, photoionization heating and collisional heating from shocks (both of which are not included in the 1D models) can keep the gas in the warm phase. Between these two regimes, the gas is marginally unstable and we discuss this below.

The multiphase nature of the ISM is much more evident in the bottom panel of Figure~\ref{fig:sim_trho}. The two trajectories in temperature-density space are immediately visible in pressure-density space. From densities of $1\ {\rm cm^{-3}}-10^{4.5}\ {\rm cm^{-3}}$, the pressure spans nearly five orders of magnitude (at fixed density the range is closer to $\sim3$~dex). For comparison, we show the observational estimates of gas pressure for SMC and LMC sight-lines from \cite{Welty2016} and they are in good agreement with our simulation.

Using these distributions, we can calculate how much mass is in each phase of the ISM. In Figure~\ref{fig:sim_mf} we show the distribution function of gas mass as a function of temperature for the stack G8 snapshots (blue) and G9 snapshots between $150-200$~Myr (orange). There are four natural breaks in the distribution that split the ISM into five phases, similar to those listed in \cite{Kim2017}. The demarcations are cold ($T<100$~K), unstable ($100~{\rm K}\leq T<5,888~{\rm K}$), warm ($5,888~{\rm K}\leq T<3.3\times10^4~{\rm K}$), ionized ($3.3\times10^4~{\rm K}\leq T<4\times10^6~{\rm K}$), and hot ($T\geq4\times10^6~{\rm K}$). For G8, 7\% of the gas is in a cold stable phase which should represent immediate fuel for star formation. A much larger fraction (57\%) exists in warm phase near $10^4$~K, consistent with \cite{Kim2017}. There is a significant amount of hotter ionized gas (18\%) that is directly related to the strength of the feedback in our simulation. This phase is key for regulating star formation as it is a significant and relatively stable gas reservoir. G9 has a much more prominent cold and unstable phase compared to G8. The ionized phase has been reduced significantly. This is partially due to the higher metallicity of G9 which allows it to cool more efficiently. However, G9 has been run for a shorter period of time than G8 so the stellar feedback has not had as long to fully develop the ionized phase. If we restrict our G8 analysis to the same time period as G9, we find that the mass in the ionized phase is reduced by a factor of two. Because there is no accretion in our isolated setup, we expect the cold/unstable phases to be depleted as time progresses and the exact distribution of mass in cosmological simulations may be different due to gas accretion. Nevertheless, the key aspect is that we predict a well developed, multiphase ISM dominated by a warm phase, in contrast to similar simulations with different physics \citep[e.g.][]{Bieri2022}.

\begin{figure}
    \centering
    \includegraphics{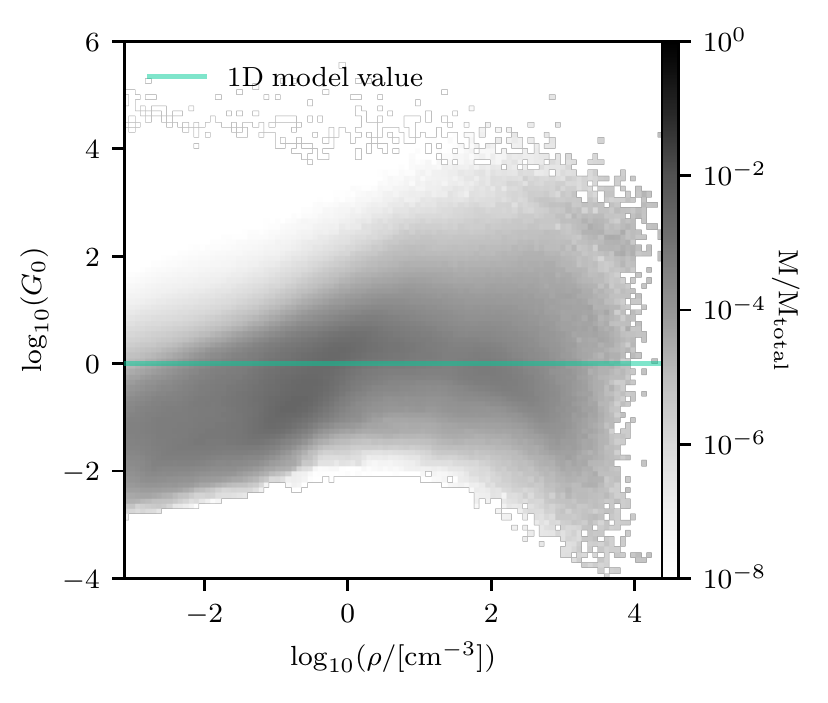}
    \caption{2D mass-weighted histogram of the strength of the FUV radiation field ($G_0$) in units of $1.6\times10^{-3}\ {\rm erg\ s^{-1}\ cm^{-2}}$ versus density for a stack of G8 snapshots from 150-400~Myr. In contrast to the 1D models (cyan), the simulated galaxy exhibits a large spread (up to eight orders of magnitude) in the strength of the FUV field at high densities which impacts the magnitude of photoelectric heating and the formation of molecules.}
    \label{fig:sim_G0rho}
\end{figure}

\subsection{The ISM radiation field}
In our 1D models, we assumed a fixed value of $G_0$, which not only sets the strength of various heating processes (e.g. photoelectric heating), but also sets the equilibrium abundances of molecules such as H$_2$ and CO as well as important ions such as C~{\small II}. However in the simulation, the local value of $G_0$ in each gas cell is set by both the radiation emitted by stars (both locally and globally) and the dust and metal\footnote{Metallic ions such as C~{\small I}, Fe~{\small I}, etc. can absorb FUV radiation.} column density.

In Figure~\ref{fig:sim_G0rho} we show a mass-weighted 2D histogram of $G_0$ versus density for the stack of G8 snapshots. At very high densities of $10^4\ {\rm cm^{-3}}$, the sub-ionizing radiation field varies by nearly eight orders of magnitude. This represents gas containing young stellar populations as well as gas that has recently cooled and condensed but has yet to form stars, and thus remains shielded. The behaviour of G9 is fundamentally similar, with even further enhanced scatter due to a higher star formation rate and higher gas densities/dust content. Towards lower densities, the scatter reduces and most of the gas has $G_0<1.0$, which is unsurprising given that the star formation rate of G8 is significantly lower than that of the Milky Way. The range of $G_0$ spanned in this histogram clearly demonstrates the necessity for on-the-fly radiative transfer. 

\begin{figure}
    \centering
    \includegraphics{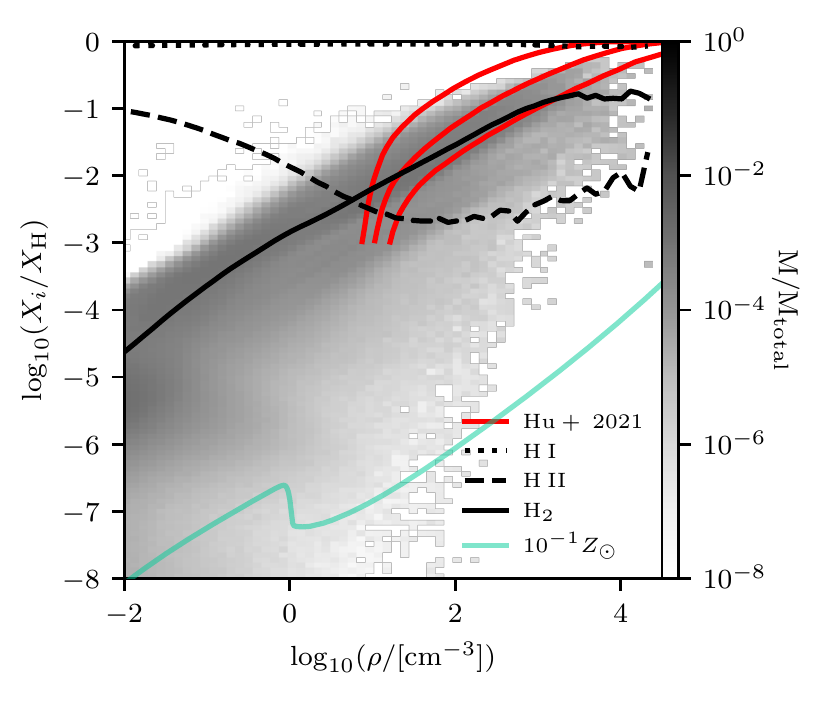}
    \includegraphics{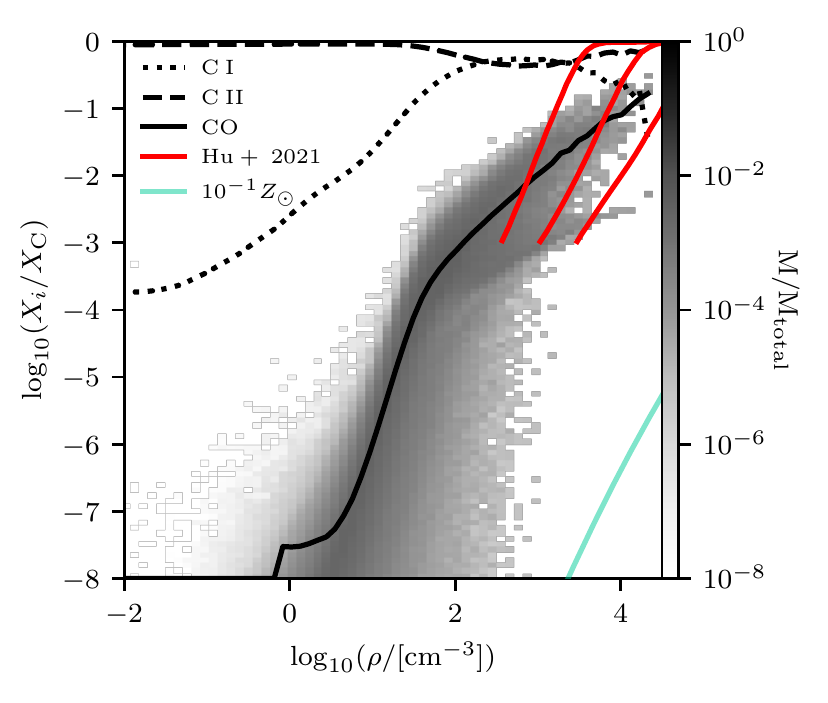}
    \caption{(Top) 2D mass-weighted histogram of the fraction of H in H$_2$ versus density for a stack of G8 snapshots from 150-400~Myr. The dashed, dotted, and solid lines show the mean fraction of H in the form of H~{\small I}, H~{\small II}, and H$_2$, respectively as a function of density. (Bottom) 2D mass-weighted histogram of the fraction of C in CO versus density for a stack of G8 snapshots from $150-400$~Myr. The various lines show the mean fraction of C in the form of C~{\small I}, C~{\small II}, and CO as a function of density as indicated in the legend. For comparison, we show the results from \protect\citealt{Hu2021} (red lines) for CO and H$_2$ at metallicities of $Z_{\odot}$, $0.3Z_{\odot}$, and $0.1Z_{\odot}$, from left to right, respectively.  In both plots, the cyan line shows the location of our 1D equilibrium model at $10^{-1}Z_{\odot}$.}
    \label{fig:sim_molrho}
\end{figure}

\begin{figure}
    \centering
    \includegraphics{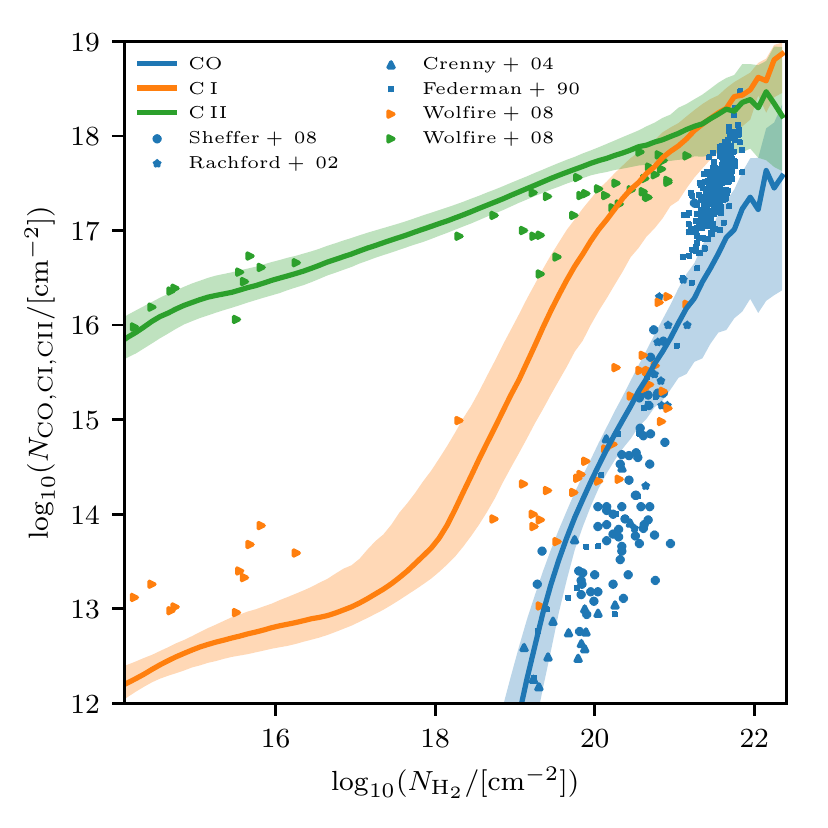}
    \includegraphics{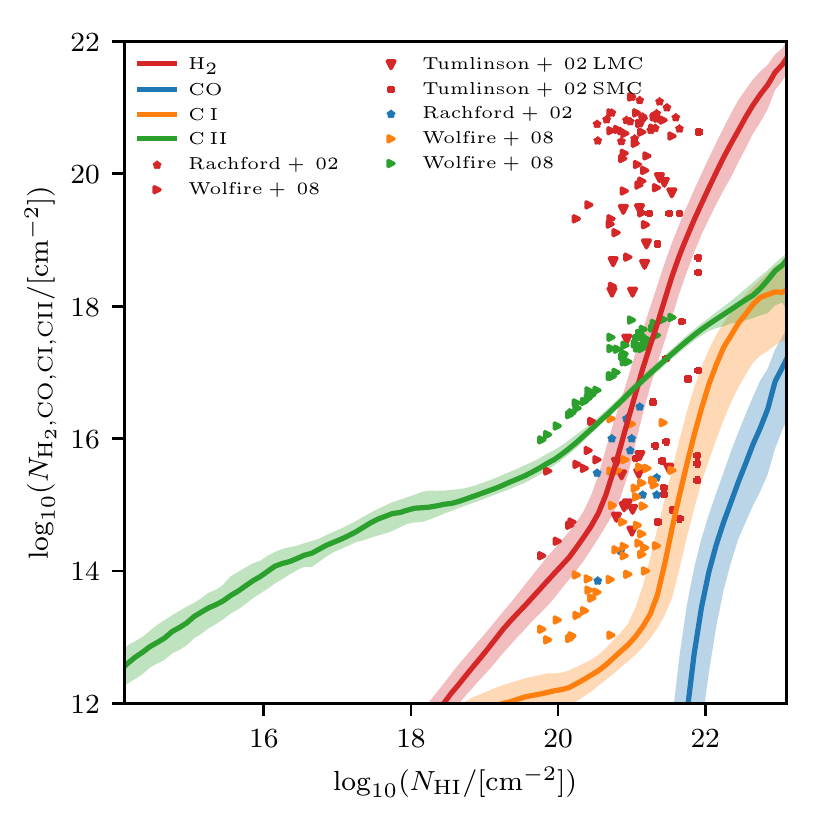}
    \caption{(Top) H$_2$ column density versus CO (blue), C~{\small I} (orange), C~{\small II} (green) column densities for a stack of G8 snapshots from 150-400~Myr. The solid line and shaded regions represent the median and 16-84 percentiles, respectively. (Bottom) H~{\small I} column density versus H$_2$ (red), CO (blue), C~{\small I} (orange), C~{\small II} (green) column densities. For comparison, we show results from galactic sight lines from \protect\cite{Federman1990,Rachford2002,Crenny2004,Wolfire2008,Sheffer2008} and LMC and SMC sight lines from \protect\cite{Tumlinson2002}.}
    \label{fig:sim_cd_comp}
\end{figure}

\subsection{The molecular and neutral content of the ISM}
Continuing with the molecular content, in Figure~\ref{fig:sim_molrho} we show 2D mass weighted histograms of the fraction of H in H$_2$ (top) and the fraction of C in CO (bottom) versus density for the stack of G8 snapshots. Compared to the 1D models, we form significantly more H$_2$ and CO. The two primary differences are the variable radiation field, which is often less than $G_0=1$ (as was assumed in the 1D models), but also the ability of the gas to self-shield. Due to the low metallicity of G8, the H$_2$ fraction never saturates. However, at $\rho\sim10^4\ {\rm cm^{-3}}$, significantly more than 10\% of H atoms are locked in H$_2$ molecules. In contrast, our solar metallicity 1D model predicted a value of $\sim10\%$ while the $0.1Z_{\odot}$ model found a value between $10^{-5}-10^{-4}$ (i.e. orders of magnitude below). We find similar behaviour in the fraction of C locked in CO atoms. At a density of $10^4\ {\rm cm^{-3}}$, the 1D model predicted a value of $\sim10^{-5}$. In contrast, 10\% of the C is in the form of CO molecules at this density in the simulation.

Comparing the top and bottom panels of Figure~\ref{fig:sim_molrho}, we find that while both the H$_2$ and CO fractions increase with density, H$_2$ has a much more gradual increase. At a density of $1~{\rm cm^{-3}}$ the H$_2$ fraction is 0.1\%. Although this is clearly subdominant, there is a substantial amount of gas in the simulation at this density. In contrast, there is essentially no CO at these same densities indicating that CO would not be a good tracer of diffuse H$_2$. A more detailed study of the relationship between H$_2$ and CO will be presented in future work.

Another noticeable difference between the 1D models and the simulation is the distribution of C~{\small I} and C~{\small II}. In the 1D models, there is no substantial amount of C~{\small I}, except in the solar metallicity model at $\rho\gtrsim10^{3}~{\rm cm^{-3}}$. There is an interesting regime in the simulation at $10^{2}~{\rm cm^{-3}}\lesssim\rho\lesssim10^{3}~{\rm cm^{-3}}$ where C~{\small I} dominates over C~{\small II}. At these densities, the radiation field on average begins to weaken while the scatter increases (see Figure~\ref{fig:sim_G0rho}) and the contribution from recombination on dust allows for a phase where C~{\small I} becomes important. Thus, we expect that fine-structure emission from C~{\small I} will trace this unique phase for galaxies like G8. C~{\small II} is the dominant phase of carbon at all other densities $>10^{-2}\ {\rm cm^{-3}}$ in this simulated dwarf galaxy.

Despite the presence of a UV background and a local radiation field, the 1D model correctly predicts the H~{\small II} fraction at a density of $10^{-2}\ {\rm cm^{-3}}$. In both the simulation and equilibrium model, the value at this density is 10\%. Towards higher densities, the H~{\small II} fraction drops off much more quickly in the 1D model while for the simulation, the mass-weighted mean H~{\small II} fraction never drops below 0.1\%. This is key because the residual electron fraction in the ISM is important for various cooling/heating processes as well as those that drive emission. At solar metallicity, the most abundant metal with an ionization potential $<13.6$~eV is carbon with a number density (with respect to hydrogen) of $2.69\times10^{-4}$ (excluding dust depletion). If the mean ionized fraction of H~{\small II} is $\gtrsim0.1\%$, this means that electrons from hydrogen atoms are a dominant contribution to the total electron fraction.

Understanding the atomic-to-molecular transition is key for modelling star formation \citep[e.g.][]{Gnedin2009,Christensen2012,Sternberg2014}. Different choices of subgrid physics, atomic data, and chemical networks will impact at what densities and how fast this transition occurs. For comparison, we show the stratified box ISM simulation predictions of the H$_2$ and CO transition of \cite{Hu2021} in Figure~\ref{fig:sim_molrho} for three different metallicities. Their 3D simulations follow a non-equilibrium network for H$_2$ and are then post-processed with carbon chemistry. We note that there are quite strong differences in both the slopes of the relations as well as the densities at which the transition onsets. This is partially due to a different reaction network and a different model for FUV field adopted here compared to \cite{Hu2021}. This comparison demonstrates the sensitivity of the transition to the chosen model.

Although our simulated galaxy is different from the Milky Way in terms of mass, metallicity, and kinematics, it is informative to compare the column densities of different ions and molecules to data from galactic sight lines where it has been measured \citep{Federman1990,Rachford2002,Crenny2004,Wolfire2008,Sheffer2008}. In Figure~\ref{fig:sim_cd_comp} we compare the median H$_2$ column densities with the CO, C~{\small I}, and C~{\small II} column densities (top) and the H~{\small I} column densities with the H$_2$, CO, C~{\small I}, and C~{\small II} column densities (bottom). Interestingly, at low $N_{\rm H_2}$, our $N_{\rm CII}$ is in very good agreement with Milky Way data but we find higher $N_{\rm CII}$ at high $N_{\rm H_2}$ compared to the Milky Way. In contrast, if we compare $N_{\rm HI}$ with $N_{\rm CII}$, our simulations underpredict the $N_{\rm CII}$ values at high $N_{\rm HI}$. Similarly we find lower $N_{\rm H_2}$, $N_{\rm CI}$, and $N_{\rm CO}$ at fixed $N_{\rm HI}$ compared to the Milky Way. We also find that G8 has higher $N_{\rm CI}$ at high $N_{\rm H_2}$ at $N_{\rm H_2}>10^{18}~{\rm cm^{-2}}$ compared to the Milky Way. This could indicate that the recombination rates on dust used in {\small PRISM} are too strong or the radiation field is much weaker which allows more C{\small I} formation. 

Compared to lower mass galaxies, G8 forms H$_2$ slightly more efficiently than the SMC but less efficiently than the LMC. This can be seen in the bottom panel of Figure~\ref{fig:sim_cd_comp} where we compare the $N_{\rm HI}$ vs. $N_{\rm H_2}$ curves to the observed data from \cite{Tumlinson2002}. 

The formation of H$_2$, C~{\small I}, and CO are all linked to dust content as H$_2$ forms on dust, CO forms via H$_2$, and C~{\small I} only becomes abundant when there is enough C enrichment, and either the dust content is high enough for recombination on dust or to shield the local FUV field. Thus the atomic-to-molecular transition in our model is likely to be sensitive to our choice of dust. The disagreement that we find when comparing various species column densities with $N_{\rm HI}$ in the Milky Way is very likely partially due to different metal and dust content of G8 compared to the Milky Way. We stress that this comparison is purely illustrative as the ISM in G8 exhibits very different conditions compared to the Milky Way. 

\begin{figure}
    \centering
    \includegraphics{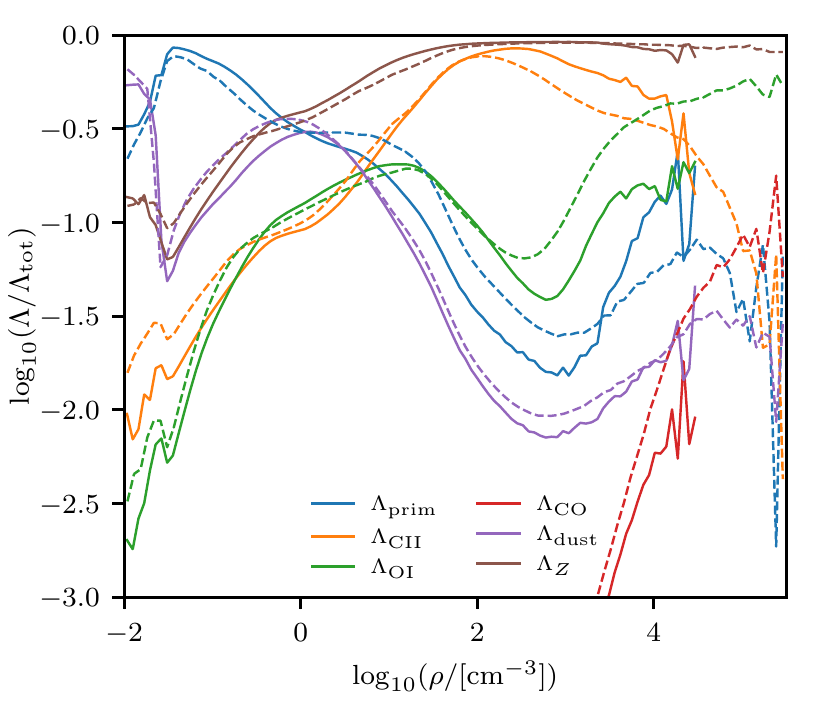}
    \includegraphics{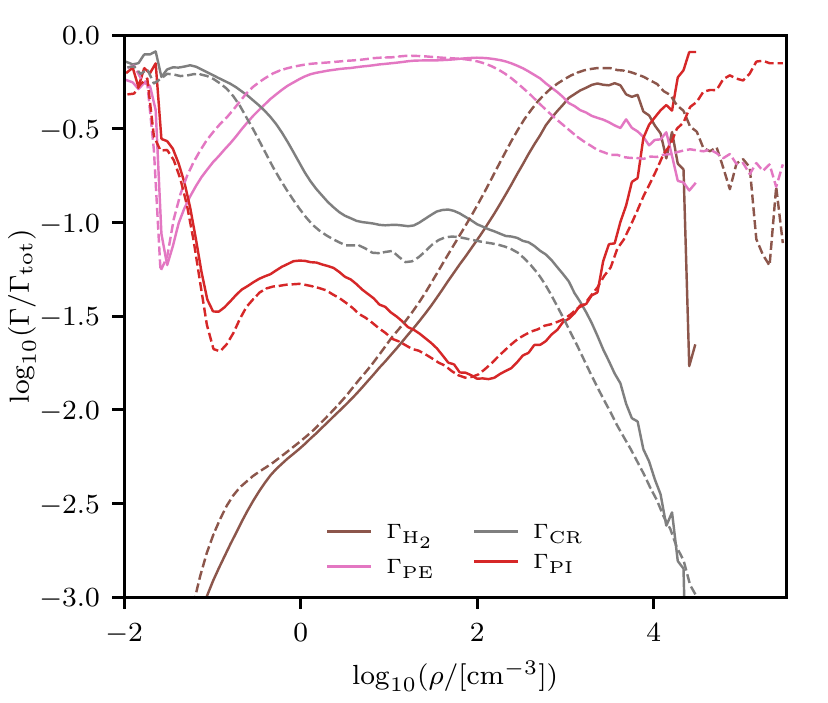}
    \caption{Fraction of the total cooling (top) or heating (bottom) contributed by various processes as a function of density as denoted in the legend. Solid and dashed lines represent a mass-weighted average for a stack of G8 snapshots from $150-400$~Myr and G9 snapshots from $150-200$~Myr, respectively.}
    \label{fig:sim_ch_rho}
\end{figure}

\begin{figure*}
    \centering
    \includegraphics[scale=1,trim={0 0cm 0cm 1.5cm},clip]{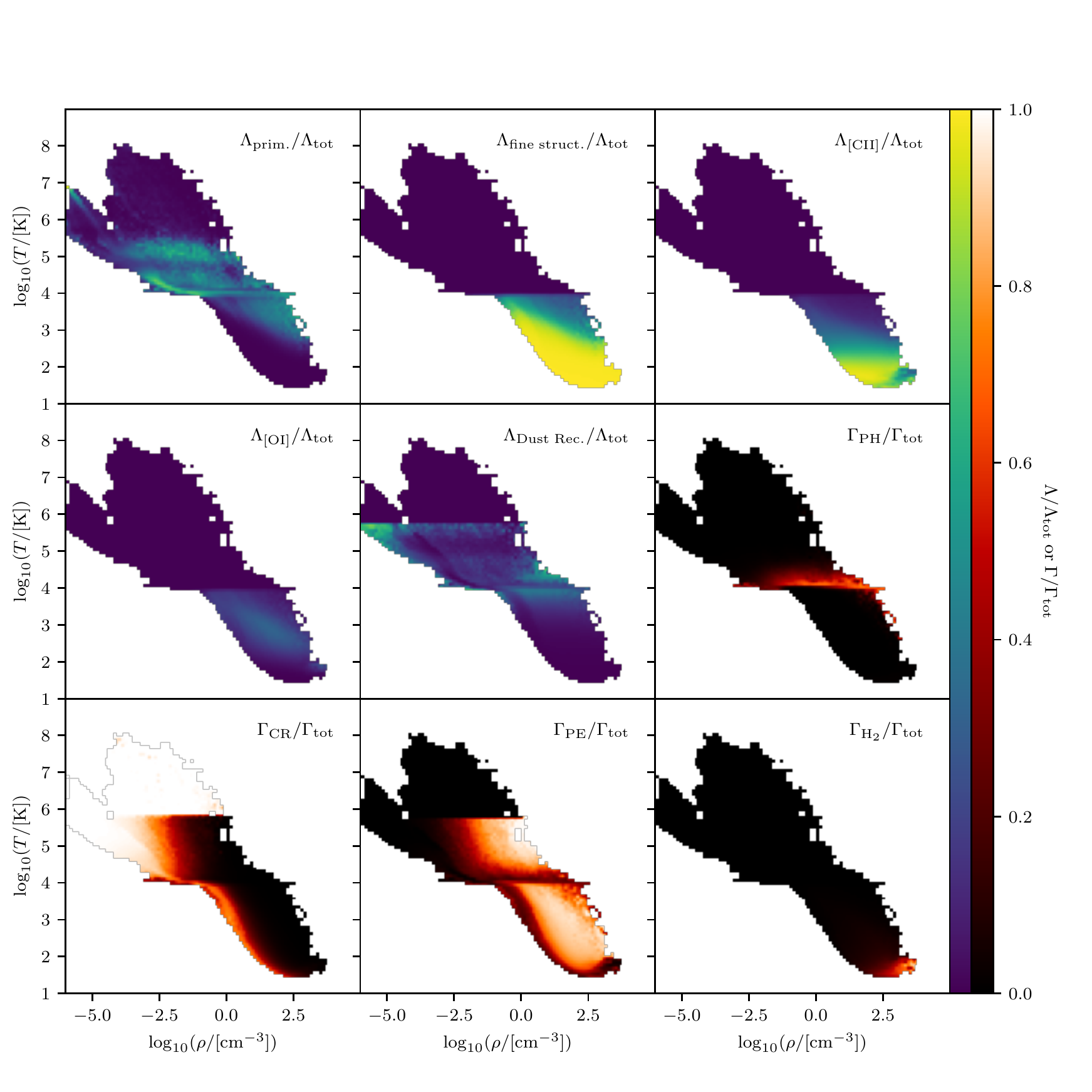}
    \caption{Density versus temperature diagrams for G8 at 400~Myr weighted by the factional contribution of various cooling and heating processes to the total. Different gas phases of the galaxy are dominated by different cooling and heating processes. Pixels are only shown if they contain at least 1~M$_{\odot}$ in gas. Note that H$_2$ cooling is included in the primordial cooling rate (top left panel). The sharp cutoff in the dust cooling and photoelectric heating are due to the assumption that all dust is destroyed at $T>10^6$~K. Furthermore, the cutoff in fine-structure cooling at $10^4$~K occurs because we switch from calculating the contribution from different ions individually assuming various collisional partners to tabulated rates assuming electrons dominate collisions.}
    \label{fig:sim_ch_trho}
\end{figure*}

\begin{figure*}
    \centering
    \includegraphics[scale=1,trim={0 1cm 0cm 1.8cm},clip]{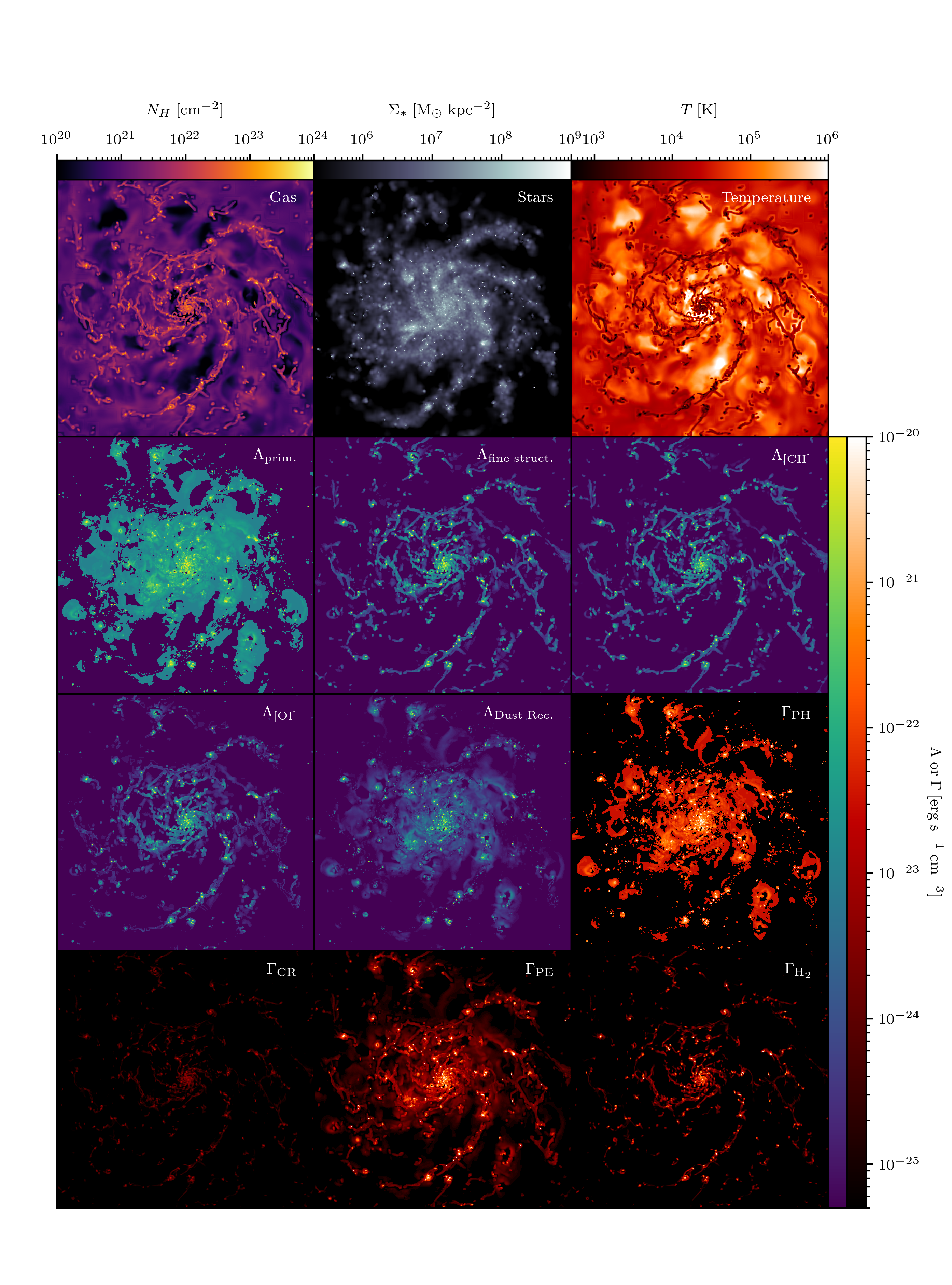}
    \caption{Maps of the total gas column density, stellar surface mass density, gas temperature (density-weighted along the line of sight), and various cooling and heating rates for different processes (density-weighted along the line of sight) for a representative snapshot of G9 at 161~Myr. The width of each image is 6.9~kpc.}
    \label{fig:sim_ch_maps}
\end{figure*}

\subsection{Which processes cool and heat the ISM?}
Constraining which processes cool and heat the ISM under a variety of conditions is of key importance for understanding how stars form. One of the primary advantages of following the individual ions and molecules fully coupled to the radiative transfer is that, in principle, our models for cooling and heating are much more self-consistent (e.g. the cooling matches the abundance and the ionization states) and accurate compared to traditionally adopted, tabulated cooling functions. 

In Figure~\ref{fig:sim_ch_rho} we show the fractional contribution of different cooling (top) and heating (bottom) processes to the total cooling or heating rate as a function of density. The values are computed as the mass-weighted average for a stack of G8 snapshots from $150-400$~Myr (solid lines) and similarly for G9 from $150-200$~Myr (dashed lines). In general, at low densities ($<0.5~{\rm cm^{-3}}$) cooling is dominated by primordial species (H, He, and H$_2$). Above this density threshold, fine-structure cooling from metal lines is the most important process. This result is very similar to the prediction from the 1D model. 

It is often the case that O~{\small I} and C~{\small II} cooling are the most important metal ions; however, the contribution from other metal are non-negligible over a wide range in density. In our model, CO cooling is a subdominant process at all densities probed. At $\rho>10^4\ {\rm cm^{-3}}$, CO contributes up to 1\% of the total cooling. This can increase in more massive, metal-enriched systems where CO may form more abundantly. Indeed we find that the contribution from CO cooling in G9 at high densities is much higher than in G8, peaking at $>10\%$. Finally, dust processes (i.e. recombination cooling and gas-grain collisional cooling) become important at a density of $\sim1\ {\rm cm^{-3}}$, almost exactly where the thermal instability allows the gas to rapidly cool (see Figure~\ref{fig:sim_trho}). Similar to the 1D model, while dust processes are rarely the most important, they contribute at least 10\% of the total cooling rate up to densities of $\sim10\ {\rm cm^{-3}}$. As we assume a relatively simplistic model for dust where the dust-to-gas-mass ratio is a fixed function of the gas-phase metallicity it is unclear how our findings will hold if dust is self-consistently formed and destroyed on-the-fly in the simulation. Future developments of the {\small PRISM} framework will allow for studying this in more detail.

In the 1D model, the transition between heating dominated by cosmic-rays versus the photoelectric effect occurred in the metallicity range $0.1Z_{\odot}-Z_{\odot}$. In the simulation, we find different behaviour. At $\rho\lesssim0.3\ {\rm cm^{-3}}$, cosmic-ray heating is generally the most important process (although there are situations where photoelectric heating and photoionization heating are also very important), as can be seen in the bottom panel of Figure~\ref{fig:sim_ch_rho}. The crossover occurs at a similar density for both G8 and G9 but the exact density at which this transition occurs may change in simulations where cosmic-rays are produced and propagated on-the-fly. Transitioning to higher densities, photoelectric heating takes over up to nearly $\rho\sim1000\ {\rm cm^{-3}}$ for G8 and at a slightly lower value for G9. At higher densities, heating due to H$_2$ formation and destruction become very important. This was not seen in the 1D models where H$_2$ formed much less efficiently due to having a fixed $G_0$ and no shielding. At $\rho\gtrsim7000\ {\rm cm^{-3}}$, the gas is mostly residing near star-forming regions and photoionization heating rapidly dominates the total heating rate. 

While Figure~\ref{fig:sim_ch_rho} highlights which cooling and heating process dominates for all mass at a given density, it is also important to recognize that variations also occur with temperature and spatially throughout the galaxy. In Figure~\ref{fig:sim_ch_trho} we show density-temperature phase-space diagrams for G8 at 400~Myr\footnote{We verified that this snapshot is representative of other points in time in the simulation.} highlighted by the mass-weighted contribution of various cooling or heating rates to the total value. While fine-structure cooling from metals is the most important cooling process at high densities, the contribution from primordial species is also highly significant in the warm, high-density gas. However, the gas is thermally unstable in these regions and thus cannot persist in this state for long. Hence, there is only a small amount of gas at high densities and high temperatures (see Figure~\ref{fig:sim_trho}).

Comparing the various heating processes, photoheating dominates a thin strip of gas at $T\sim10^4$~K as well as the highest density and highest temperature gas. There is clear separation between where cosmic-ray, photoelectric, and H$_2$ heating are most important. Photoelectric and cosmic-ray heating both operate at all densities and temperatures, but cosmic-ray heating is more important at the lowest densities at fixed temperature, while the opposite is true for photoelectric heating. H$_2$ heating is only relevant at the highest densities and lowest temperatures where it forms most efficiently. 

We note two particular features in Figure~\ref{fig:sim_ch_trho}. The metal fine-structure contribution appears to become negligible above $10^4$~K and heating and cooling associated with dust goes to zero above $10^6$~K. In practice, we compute metal cooling at all temperatures; however, at $T\gtrsim10^4$~K, metal cooling is dominated by collisional processes with electrons, while at lower temperatures other collisional partners matter. Here we show only the contribution at $T<10^4$~K\footnote{The two regimes are separated in the code as the former can easily be tabulated. }. Furthermore, in our simplistic dust model, we assume that dust cannot exist at $T>10^6$~K. 

We also emphasize that here we only show heating or cooling processes that are explicitly calculated in our ISM model. Processes such as the $PdV$ (adiabatic cooling/heating) work from SN explosions, shock heating, and gravitational compression heating, while calculated in the code, are not counted in the total cooling or heating rates shown. This is because these processes are operator split from the cooling and heating in {\small PRISM} and occur in the internal energy update when solving the Euler equations. Nevertheless, their importance should not be understated as they are key for driving gas to very high temperatures. 

To conclude our analysis on which processes heat and cool the ISM, we show in Figure~\ref{fig:sim_ch_maps} density-weighted maps of various cooling and heating rates for G9 at 168~Myr\footnote{The results are similar for G8.}. For comparison, we also show maps of the total gas column density, stellar surface mass density, and gas temperature. Beginning with cooling, fine-structure metal-line cooling is primarily concentrated in the dense gas filaments. Cooling from primordial species, in contrast, extends into the much more diffuse gas between filaments. Dust recombination cooling is more concentrated around star-forming regions that are actively being metal enriched.

Continuing with heating processes, photoheating is patchy and traces the youngest star-forming regions. In some locations it is very concentrated around young stars when the nebula is ionization-bounded. However when stars are located in density-bounded gas clouds, the escape fraction from the birth cloud is high and photoheating can extend well beyond the star-forming region. Unsurprisingly, H$_2$ heating traces the densest gas filaments and is much more concentrated than, for example, [C~{\small II}] or [O~{\small I}] cooling. This has important implications for how well fine-structure emission from either of these two metals trace H$_2$. Cosmic-ray heating is clearly the weakest of all processes shown so it is difficult to visually assess its importance in comparison to, e.g. photoelectric heating. This is primarily because it is most important in diffuse gas which, in general, has a low heating rate compared to other regions in the galaxy. 

In summary, one of the primary advantages of the {\small PRISM} model is being able to take a census, fully in non-equilibrium, of where each cooling and heating process is regulating the thermodynamics of the ISM. It is clear that the contribution of each process is more complex in a dynamic 3D system than the 1D equilibrium models suggest. Below we will continue by assessing the importance of calculating cooling and heating in non-equilibrium.

\subsection{When do non-equilibrium effects matter in the ISM?}
One of the primary advantages of the coupling between the {\small PRISM} model and {\small RAMSES-RTZ} is that we can explore the impact of non-equilibrium effects on the properties of the ISM. If the gas is significantly out of equilibrium, the 1D models will not be able to make robust predictions for ISM properties and observational signatures, which may bias our understanding of galaxy formation physics. In this Section, we identify where the gas is in and out of equilibrium and why.

\subsubsection{Thermal Disequilibrium}
\begin{figure}
    \centering
    \includegraphics[scale=1,trim={0 0.3cm 0cm 1.1cm},clip]{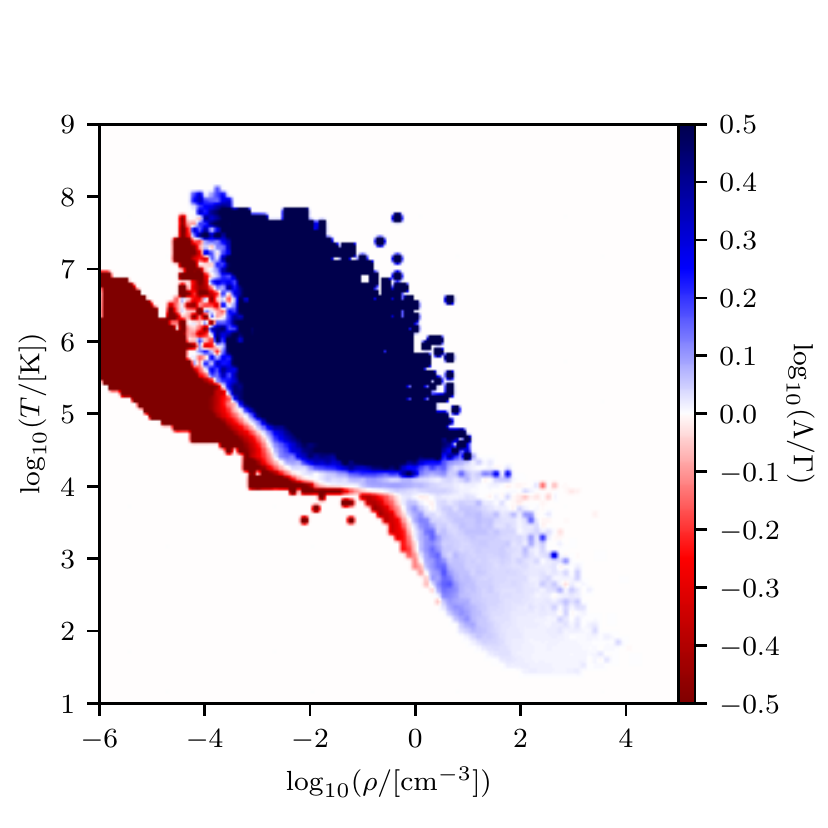}
    \includegraphics[scale=1,trim={0 0.3cm 0cm 1.1cm},clip]{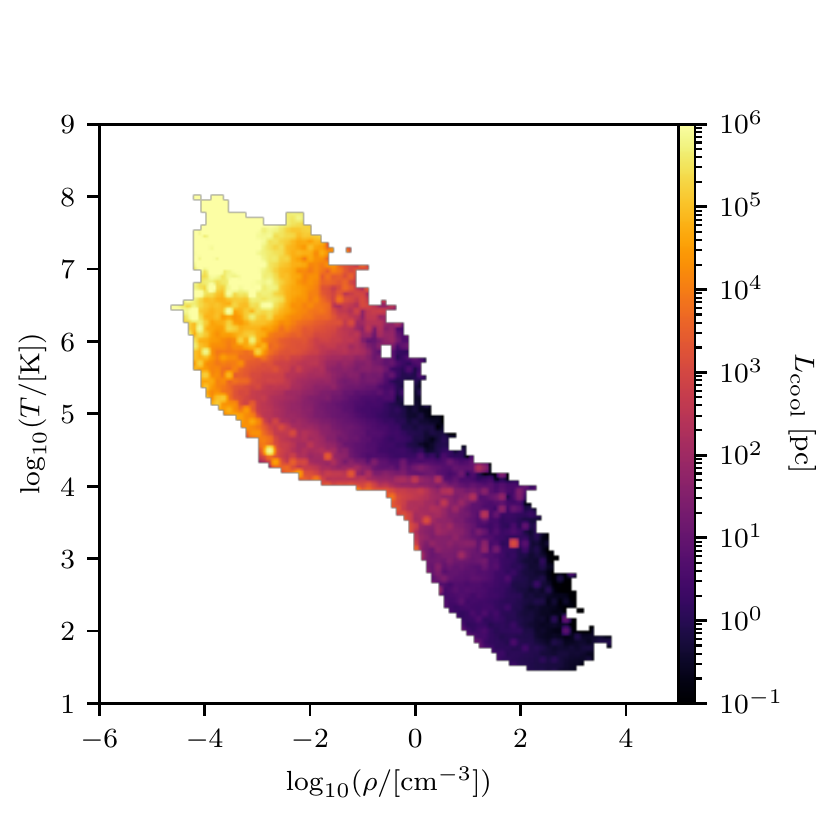}
    \includegraphics[scale=1,trim={0 0.3cm 0cm 1.1cm},clip]{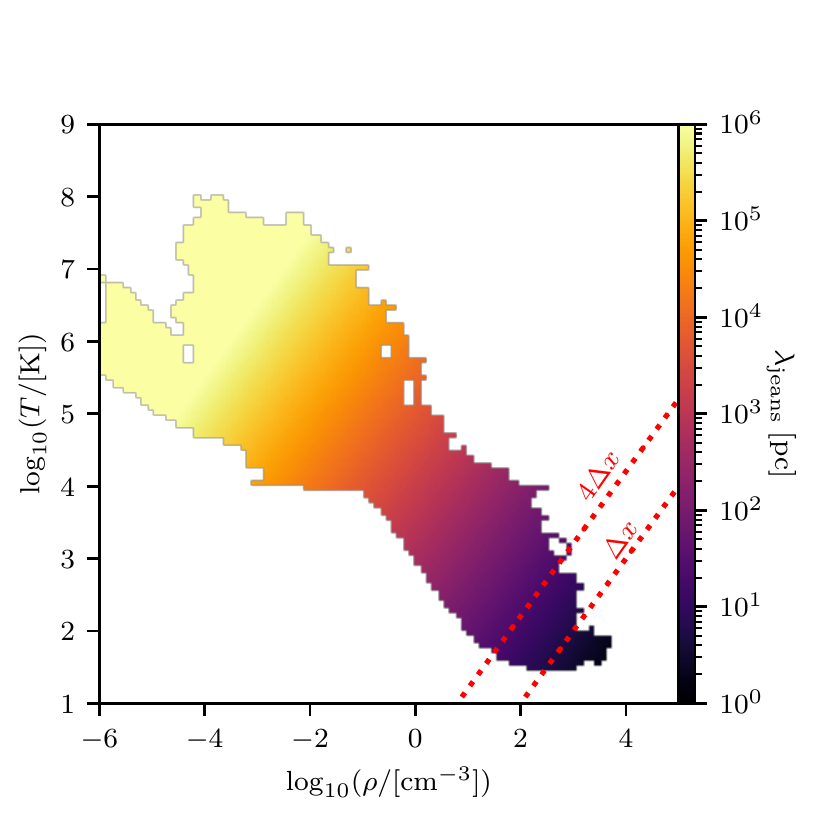}
    \caption{(Top) Logarithm of the ratio between the mass-weighted cooling and heating rates for G8 at $400$~Myr as a function of temperature and density. Blue and red regions demonstrate where cooling or heating dominate, respectively, while white regions are in thermal equilibrium. (Middle) Mass-weighted cooling length for thermally unstable, cooling gas in G8 at $400$~Myr as a function of temperature and density. (Bottom) Jeans length ($\lambda_{\rm jeans}$) for gas cells in G8 at 400~Myr. The dotted red lines show contours of constant $\lambda_{\rm jeans}$ at the maximum resolution and four times the maximum resolution in our simulation.}
    \label{fig:therm_neq}
\end{figure}

We begin by analyzing where the gas is in and out of thermal equilibrium. In the top panel of Figure~\ref{fig:therm_neq} we show the log ratio of total cooling to heating rate for G8 at $400$~Myr. White regions represent where the gas is close to thermal equilibrium whereas blue and red regions indicate where cooling or heating dominate, respectively. There is a thin white band that extends from $T\sim10^8$~K all the way down to $T\lesssim10^4$~K where it bifurcates. The stronger branch extending towards lower temperatures follows the standard equilibrium cooling curve that one would obtain from collisional ionization equilibrium. This is similar to the cooling curves shown in Figure~\ref{fig:trho}. The horizontal branch includes a photoheating contribution from star particles that maintains the ionization state and high temperature. These trajectories are exactly the same as those seen in the mass-weighted temperature-density diagram (Figure~\ref{fig:sim_trho}).

In general, the dense ISM is either very close to thermal equilibrium or has a slightly higher cooling than heating rate, indicative of a cooling instability. There are a few red pixels at high densities, which are regions around young star particles where photoheating is actively heating the gas. The lowest density regions of the ISM at $T<10^4$~K are also red, partially from the impact of cosmic-ray heating. According to our model, the gas with the highest cooling to heating ratio is that at $T>10^4$~K which primarily resides outside of the disk (although there is a warm and hot component of the disk as well). This is the gas that has been recently heated by SN feedback and driven out of the galaxy in a wind. Thermal instability in galactic winds within the context of our model is further studied in Rey~et~al.~{\it in prep.}

One of the important consequences of the cooling instability is that it can lead to gas fragmentation on the scale of the cooling length \citep[e.g.][]{McCourt2018,Sparre2019,Mandelker2021}, defined as $L_{\rm cool}=c_{\rm s}t_{\rm cool}$, with $c_{\rm s}$ being the sound speed and $t_{\rm cool}$ being the cooling timescale. Following this line of reasoning, we can estimate what length scale we need to resolve in the simulation in order to capture this fragmentation (see also \citealt{Smith2017}). In the middle panel of Figure~\ref{fig:therm_neq} we show the mass-weighted cooling length as a function of density and temperature for thermally unstable, cooling gas for G8 at $400$~Myr. The densest regions of the ISM that are actively cooling have cooling lengths $\ll 1$~pc, far below the resolution scale of our simulation. Many of these same cells also have a Jeans' length below the resolution limit of the simulation (see the bottom panel of Figure~\ref{fig:therm_neq}). It is important to consider that 58\% of the gas by mass is in thermal equilibrium or actively heating; however, among the cooling gas, 33\% by mass has a cooling length smaller than the finest resolution element in our simulation. Although unsurprising, this indicates that the simulation is far away from structural convergence\footnote{We note that this analysis does not account for non-thermal pressure support, such as that from turbulence, so the exact details of the fragmentation are subject to additional physics beyond a cooling or Jeans' instability.} (see also \citealt{Hummels2019,Peeples2019} for a similar discussion in the CGM). In other words, there could be structure below the resolution scale of the simulation that is important for chemistry, heating, cooling, and various other physical processes. Therefore, future versions of {\small PRISM} may require subgrid modelling for the density structure below the scale of the smallest resolution element to converge predictions with increasing resolution \citep[e.g.][]{Liu2022,Buck2022}.

\subsubsection{Chemical Disequilibrium}
It is possible that the gas can be close to thermal equilibrium while being out of chemical (or ionization) equilibrium. It is well established that in the ISM, cooling timescales can be faster than recombination timescales, which leads to an ionization lag, where the gas is over-ionized compared to equilibrium \citep[e.g.][]{Kafatos1973,Gnat2007,Oppenheimer2013,Vasiliev2013}. Alternatively, if the heating timescale is faster than the ionization timescale, the gas may be under-ionized compared to equilibrium \citep[e.g.][]{Klessen2016}. Such an effect can be realized near shocks. In this Section, we discuss where the gas is out of chemical equilibrium and focus particularly on the electron fraction as it controls a substantial amount of heating and cooling processes as well as line emission from the galaxy.

\begin{figure}
    \centering
    \includegraphics[scale=1,trim={0 0cm 0cm 1.1cm},clip]{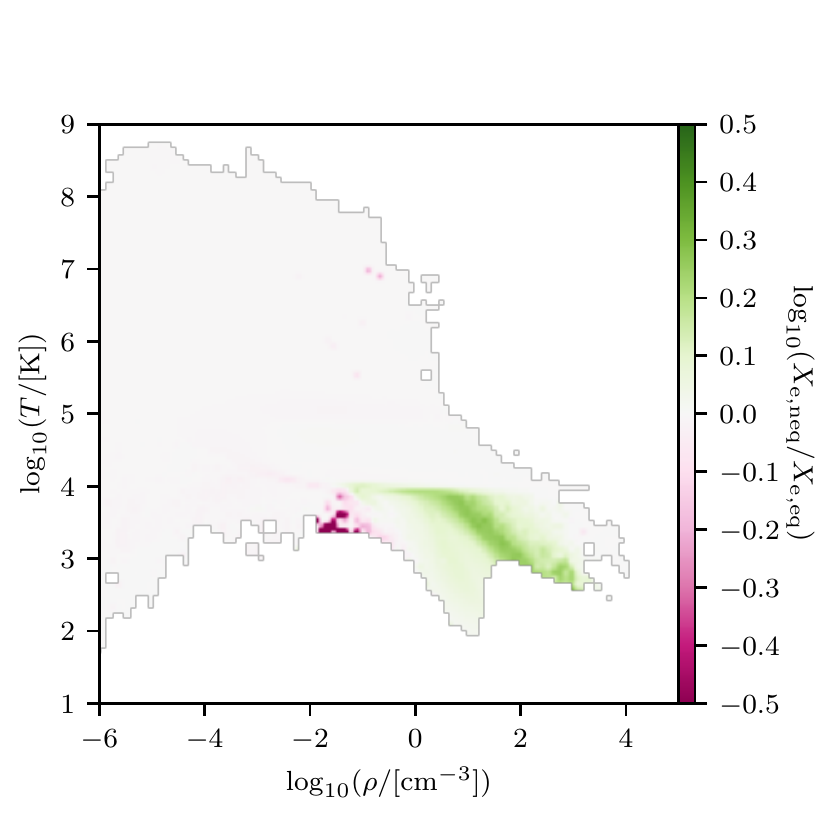}
    \caption{Logarithm of the ratio between the mass-weighted non-equilibrium and equilibrium electron fractions as a function of density and temperature in the G8 snapshot at 175~Myr. Gas is only shown if the electron fraction is at least 0.1\%.}
    \label{fig:e_neq}
\end{figure}

To calculate whether a gas cell is in chemical equilibrium, we restart G8 from all snapshots between $175-250$~Myr, fixing the temperature and radiation field, and evolve the simulation until chemical equilibrium is reached. In Figure~\ref{fig:e_neq} we show a 2D histogram of density versus temperature coloured by the log ratio of non-equilibrium to equilibrium mass-weighted electron fraction for a stack of all snapshots in this time period. Gas is only shown if it has an electron fraction $>0.1\%$ in either the equilibrium or non-equilibrium output. In certain cases, the non-equilibrium electron fraction is more than three times the equilibrium value. Similarly high ratios were observed in \cite{Richings2016}. We believe that the origin of this behaviour is due to the same cooling instability that has been previously discussed in the literature \citep[e.g.][]{Kafatos1973,Gnat2007,Oppenheimer2013,Vasiliev2013}, i.e. the gas cooling rate is faster than the recombination rate. 

Given the Eulerian nature of the code, we cannot currently track the temperature and ionization history of individual gas parcels in the simulation\footnote{This will be remedied in future iterations of the {\small PRISM} model coupled to Monte Carlo tracer particles \citep{Cadiou2019}.}. It is important to note that at the current time, the cooling timescale does not need to be greater than the ionization timescale, rather, for an ionization lag to manifest, the cooling instability could have occurred any time in the very recent past. 

We verified that the H~{\small II} distribution exhibits the same out-of-equilibrium properties as the electron distribution, confirming that the source of the excess electrons are from H atoms. There are not enough metals for charge exchange reactions to push H~{\small II} this far out of chemical equilibrium, ergo we conclude that catastrophic cooling is likely responsible for the gas that has strong non-equilibrium electron enhancements.

Figure~\ref{fig:e_neq} also shows that some gas has a non-equilibrium electron deficit compared to the equilibrium value (see also \citealt{Richings2016}). In these regions, the gas has been heated faster than it can be ionized. This gas tends to occupy much lower densities than the gas that is over-ionized compared to equilibrium. Strong shocks have been proposed as one mechanism to heat gas faster than it can be collisionally ionized \citep[e.g.][]{Klessen2016}. In our simulations, SN feedback in dense regions could be the source of these shocks. However, here we propose that cosmic-ray heating might also cause this effect.

\begin{figure}
    \centering
    \includegraphics[scale=1,trim={0 0cm 0cm 0.4cm},clip]{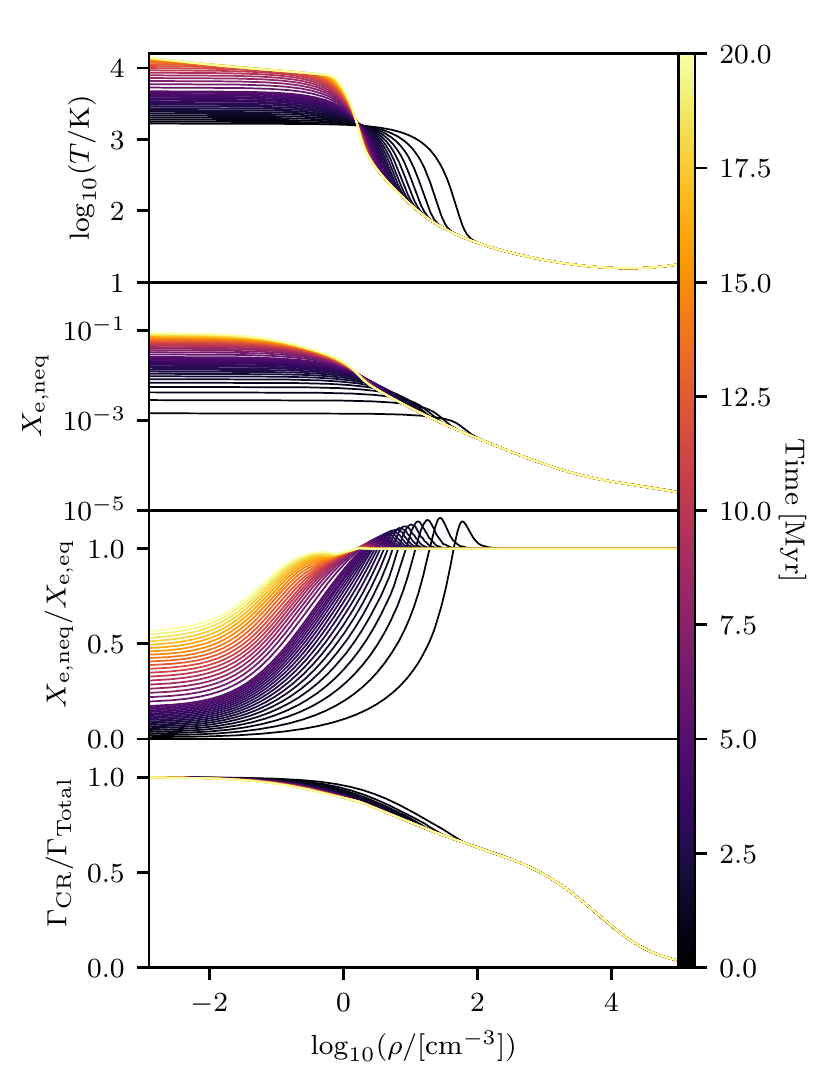}
    \caption{Gas temperature (first panel), non-equilibrium electron fraction (second panel), ratio of non-equilibrium to equilibrium electron fraction (third row), and ratio of cosmic-ray heating rate to the total heating rate (bottom panel) as a function of gas density for our fiducial 1D model at $10^{-1}Z_{\odot}$ with $G_0=1$ $\eta_{\rm cr}=10^{-16}\ {\rm s^{-1}\ H^{-1}}$. Different colours represent different times as shown in the colour bar.}
    \label{fig:xe_eq_1d}
\end{figure}

To demonstrate this, we have rerun our fiducial 1D model at $10^{-1}Z_{\odot}$ with $G_0=1$ and $\eta_{\rm cr}=10^{-16}\ {\rm s^{-1}\ H^{-1}}$ generating a series of outputs between 0 and 20~Myr. In these simulations, we have initialized the gas temperature to 3~K. For each output, we then restart the calculation, fixing the temperature, and evolving the ionization states to equilibrium. In Figure~\ref{fig:xe_eq_1d} we show the gas temperature, non-equilibrium electron fraction, and ratio of non-equilibrium to equilibrium electron fraction as a function of density and time. At densities $>10^2\ {\rm cm^{-3}}$ the system is in chemical and thermal equilibrium beginning at timescales $\ll 1$~Myr. However, at low densities (i.e. $\rho<10^{-1}\ {\rm cm^{-3}}$), the gas is significantly under-ionized, even at 20~Myr, when it is in thermal equilibrium. As we have shown in Figure~\ref{fig:sim_ch_rho}, cosmic-ray heating dominates the heating rate at these densities in our simulation. We further demonstrate that this is the case in the 1D tests by showing the ratio of $\Gamma_{\rm CR}/\Gamma_{\rm Total}$ as a function of density and time. Indeed at these low densities, cosmic-rays account for nearly 100\% of the total heating rate.

At intermediate densities (i.e. $10^0<\rho/{\rm cm^{-3}}<10^2$), there is a transient effect where the gas can be over-ionized compared to equilibrium by $\sim10\%$. This is not enough to account for the over-ionization observed in Figure~\ref{fig:e_neq}, which exhibits values upwards of a factor of three. The origin of this effect also seems to be that cooling is faster than recombination. In the top row of Figure~\ref{fig:xe_eq_1d}, at early times, the temperature over-shoots the equilibrium value. This over-shoot allows the gas to be ionized above the equilibrium value (see second and third rows of Figure~\ref{fig:xe_eq_1d}). As the gas cools down, the cooling timescale becomes shorter than the recombination timescale allowing for a small amount of over-ionization. 

In summary, we have shown that there are regions of the ISM where, by mass, the electron fraction is both enhanced and reduced compared to equilibrium. This impacts not only the cooling and heating in the ISM, but also observable emission \citep{Richings2022,RTZ}. In future work, we will explore how star formation and line emission change due to non-equilibrium effects and more generally the {\small PRISM} model.

\section{Known Limitations and Future Developments}
\label{sec:limits}
Before proceeding to our conclusions, we highlight a few important limitations. As described above, we assume a dust-to-gas mass ratio based on oxygen abundance and a fixed dust composition. This assumption is unlikely to reflect the dust composition and abundance in all galactic environments and in some cases does not explicitly conserve mass. Furthermore, we apply a simplistic model for dust temperature. Future work that models dust production and destruction coupled to the radiation field in the context of {\small RAMSES-RTZ} and the {\small PRISM} model will ideally remedy these issues (Rodriguez~Montero~et~al.~{\it in prep}).

Thermal conduction is a potential important source of heat transport \citep{Spitzer1962} that is not included in our fiducial model. An accurate calculation of thermal conduction relies on properly modelling magnetic fields (which themselves can be an important pressure term in the ISM). We leave the inclusion of thermal conduction and magnetic fields to future work.

Our uniform model for cosmic-rays is unlikely to be an adequate representation of how cosmic-rays are actually distributed throughout galaxies. Current generations of simulations can now include cosmic-ray production, streaming, and diffusion, self-consistently \citep[e.g.][]{Chan2019,Farcy2022}, and we anticipate a more accurate cosmic-ray implementation being an important future development for our model.

The cooling processes presented in this work are often only accurate in the optically thin limit. This is especially true for our models of metal-line cooling and the exact density where the gas becomes optically thick depends on the specific transition. This can result in inaccuracies when the density of the gas goes above the critical density of the transition. For this reason, we recommend caution in applying our model in regimes that are not in the optically thin limit. Fortunately, densities of this magnitude are not typically reached by modern cosmological simulations; however, this may become problematic in giant molecular cloud simulations with sub-pc resolution aimed at modelling star formation. 

Due to finite spatial and mass resolution, there is an upper limit on density that can be reached with simulations. This can impact the chemistry due to the $n^2$-dependence of most of our chemical reactions. When important density structures are unresolved, our model may under-produce molecules such as CO and H$_2$ that tend to form at high densities. For this reason, we have adopted a sub-grid clumping factor, $C$, for H$_2$ formation, which impacts CO since CO formation is dependent on H$_2$ number density. Our fiducial model has $C=1$, but depending on the simulation, this may need to be calibrated \citep{Gnedin2009}, e.g. on the observed H$_2$ column density distribution in the LMC or SMC \citep[e.g.][]{Tumlinson2002}.

\section{Conclusions}
\label{conclusion}
We have introduced the {\small PRISM} ISM model for thermochemistry and its coupling to the {\small RAMSES-RTZ} code. {\small PRISM} accounts for the dominant cooling and heating processes (i.e. photoheating, photoelectric heating, H$_2$ heating/cooling, cosmic-ray heating, primordial species cooling, metal-line cooling, CO cooling, and dust cooling) in the low-density ISM (i.e. $\rho\lesssim10^5\ {\rm cm^{-3}}$) as well as a non-equilibrium chemical network of up to 115 different species including the H$_2$ and CO molecules. All of these physical processes are fully coupled to the on-the-fly multifrequency radiation transport available in the {\small RAMSES} code \citep{Teyssier2002,Rosdahl2013}. The combination of {\small RAMSES-RTZ} and {\small PRISM} is among the first suites of software to couple such a large chemical network to on-the-fly radiation hydrodynamics in an efficient enough manner to run full 3D cosmological \citep{Katz2022-popiii} and isolated galaxy simulations \citep{Cameron2022}. 

We have validated that {\small PRISM} model across six decades in metallicity by comparing with 1D equilibrium models in the literature \citep{Koyama2000,Wolfire2003,Bialy2019,JKim2022}. We then applied the model to two isolated dwarf galaxy simulations (G8 and G9, similar in many respects to the SMC and LMC, \citealt{Rosdahl2015}) to take a census of which cooling and heating processes dominate different regions of a galaxy and to assess the importance of non-equilibrium effects. The 3D simulations differ in many ways from the 1D models due to the inclusion of ionizing radiation, a variable FUV radiation field, and self-shielding. This manifests in a much faster atomic-to-molecular transition in the simulations compared to the 1D models. Furthermore, we show that the electron fraction in the ISM can be more than three times enhanced or reduced in the ISM because of non-equilibrium effects (see also \citealt{Richings2016}). We attribute the enhancement to recombination lags where the cooling rate is faster than the recombination timescale. In contrast, we attribute electron fraction deficits to rapid cosmic-ray heating (in contrast to strong shocks), which is behaviour that can be reproduced in time-dependent 1D models.

Our presented model has known limitations. For example, we employ an effective model for dust, ignore magnetic fields and thermal conduction, do not self-consistently track cosmic-ray production and transport, and the density squared dependence of many of the reactions leads to a resolution dependence in their abundance. In future work, we will develop the {\small PRISM} model to address these limitations in an effort to more accurately model the ISM and understand galaxy formation. The model is suitable for addressing numerous problems related to galaxy formation physics and the ISM. As a first application, it has recently been used to demonstrate the impact of temperature fluctuations of metallicity measurements \citep{Cameron2022}. Moving forward, we will extend our simulation suite to diverse environments from galaxy formation in a cosmological context to simulations of individual molecular clouds.

\section*{Acknowledgements}
HK thanks Romain Teyssier for making the {\small RAMSES} code public. TK was supported by the National Research Foundation of Korea (NRF) grant funded by the Korea government (No. 2020R1C1C1007079 and No. 2022R1A6A1A03053472). MR is supported by the Beecroft Fellowship funded by Adrian Beecroft. OA and EA acknowledge financial support from the Knut and Alice Wallenberg Foundation and the Swedish Research Council (grant 2019-04659). This work was performed using the DiRAC Data Intensive service at Leicester, operated by the University of Leicester IT Services, which forms part of the STFC DiRAC HPC Facility. The equipment was funded by BEIS capital funding via STFC capital grants ST/K000373/1 and ST/R002363/1 and STFC DiRAC Operations grant ST/R001014/1. DiRAC is part of the National e-Infrastructure.

\section*{Data Availability}
The data underlying this article will be shared on reasonable request to the corresponding author.

\appendix
\section{Equilibrium Temperature-Density Comparison of {\small PRISM} with \protect\cite{RTZ} physics}
\label{app:changes}
\begin{figure}
    \centering
    \includegraphics{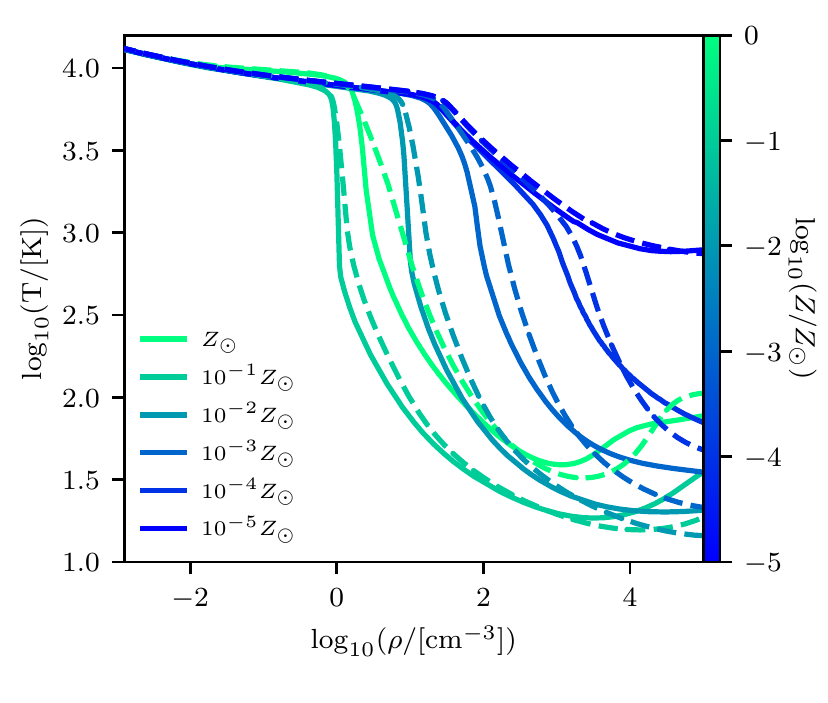}
    \caption{Equilibrium temperature-density relations as a function of metallicity. The solid lines show the fiducial {\small PRISM} model while the dashed lines represent the model using our previous fine-structure cooling collisional rates and not assuming the factor 1.5 boost in the photoelectric heating, as was done in \protect\cite{RTZ}. There are minimal differences between the two models with the previous model having slightly less efficient cooling.}
    \label{fig:trho_comp}
\end{figure}

In Figure~\ref{fig:trho_comp} we compare the equilibrium temperature-density curves at various metallicities when using the fine-structure cooling collisional rates from \protect\cite{RTZ} and not assuming the factor 1.5 increase in the photoelectric heating due to the updated PAH abundance as was done for our simulations. G8 and G9 were initialized at $0.1Z_{\odot}$ and $0.5Z_{\odot}$ and we see only small differences between the different models at $\rho\lesssim10^3\ {\rm cm^{-3}}$ for these metallicities. There is some difference at higher densities, primarily due to a change in the O~{\small I} cooling. Our new model tabulates effective collision rates directly from {\small CLOUDY} \citep{Ferland2017} for all fine-structure transitions in our model for the following collision partners: H, H$^+$, ortho and para H$_2$, $e^-$, He, He$^+$, and He$^{++}$.

\bibliographystyle{mnras}
\bibliography{example} 

\bsp	
\label{lastpage}
\end{document}